\documentclass[reqno]{amsart}
\usepackage[utf8]{inputenc}

\usepackage[citecolor=blue]{hyperref}

\usepackage[backend=biber, 
			style=authoryear, 
			natbib,  
			maxcitenames=2,
			maxbibnames=99, 
			uniquename=false]{biblatex}
\usepackage[utf8]{inputenc} 
\usepackage{graphicx}
\usepackage{nicefrac}       
\usepackage{microtype}      
\usepackage{amsmath, amsfonts,bm,amssymb,indentfirst,mathtools,bbm}
\usepackage[boxed]{algorithm2e}
\SetAlCapSkip{1em}
\usepackage{color}
\usepackage[title]{appendix}
\usepackage[shortlabels,inline]{enumitem}
\usepackage{multirow}

\usepackage[frozencache,cachedir=.]{minted}

\allowdisplaybreaks[4]

\numberwithin{figure}{section}
\numberwithin{equation}{section}

\makeatletter
\newcommand*{\rom}[1]{\expandafter\@slowromancap\romannumeral #1@}
\newcommand\niton{\mathrel{\m@th\mathpalette\canc@l\owns}}
\newcommand\canc@l[2]{{\ooalign{$\hfil#1/\mkern1mu\hfil$\crcr$#1#2$}}}
\makeatother

\usepackage{fancyhdr}
\usepackage{hyperref}
\hypersetup{colorlinks=true}

\usepackage[matrixnorms,opnorm=op]{arash_macros}
\usepackage{xcolor}

\DeclareMathOperator{\Beta}{\mathsf{Beta}}

\DeclareMathOperator{\unif}{\mathsf{Unif}}

\DeclareMathOperator{\gem}{\mathsf{GEM}}
\DeclareMathOperator{\sbm}{\mathsf{SBM}}
\DeclareMathOperator\DP{\mathsf{DP}}

\DeclareMathOperator{\bern}{\mathsf{Ber}}

\newcommand\ind[1]{1_{\{#1\}}}
\newcommand\pib{\bm \pi}
\newcommand\ub{\bm u}
\newcommand\vb{\bm v}
\newcommand\Ab{\bm A}
\newcommand\yb{\bm y}
\newcommand\thetab{\bm \theta}

\newcommand\zb{\bm z}
\newcommand\wb{\bm w}
\newcommand\etab{\bm \eta}
\newcommand\xib{\bm \xi}

\newcommand\given{\,|\,}
\newcommand{\m}{m}

\newcommand\n{n}
\newcommand\thb{\bm \theta}

\usepackage[foot]{amsaddr}

\usepackage{float}

\usepackage[font=small,skip=0pt]{caption}
\usepackage{adjustbox}

\usepackage{booktabs}
\usepackage{array}
\newcolumntype{P}[1]{>{\centering\arraybackslash}p{#1}}

\title[NSBM for clustering]{Nested stochastic block model for simultaneously clustering networks and nodes}

\author{Nathaniel Josephs$^1$}
\author{Arash A. Amini$^2$}
\author{Marina Paez$^3$}
\author{Lizhen Lin$^4$}

\address[1]{Department of Statistics, North Carolina State University}
\address[2]{Department of Statistics, UCLA}
\address[3]{Department of Statistical Methods, Federal University of Rio de Janeiro}
\address[4]{Department of Mathematics, The University of Maryland}

\begin{document}

\maketitle

\vspace{-2em}

\begin{abstract}
    We introduce the nested stochastic block model (NSBM) to cluster a collection of networks while simultaneously detecting communities within each network.
    NSBM has several appealing features including the ability to work on unlabeled networks with potentially different node sets, the flexibility to model heterogeneous communities, and the means to automatically select the number of classes for the networks and the number of communities within each network.
    This is accomplished via a Bayesian model, with a novel application 
    of the nested Dirichlet process (NDP) as a prior to jointly model the between-network and within-network clusters.
    The dependency introduced by the network data creates nontrivial challenges
    for the NDP, especially in the development of efficient samplers.
    For  posterior inference, we propose several Markov chain Monte Carlo algorithms including a standard Gibbs sampler, a collapsed Gibbs sampler, and two blocked Gibbs samplers that ultimately return two levels of clustering labels from both within and across the networks.
    Extensive simulation studies are carried out which demonstrate that the model provides very accurate estimates of both levels of the clustering structure.
    We also apply our model to two social network datasets that cannot be analyzed using any previous method in the literature due to the anonymity of the nodes and the varying number of nodes in each network.

\end{abstract}

\section{Introduction}

Suppose we have a collection of networks represented by a sequence of adjacency matrices.
How do we simultaneously cluster these networks and the nodes within?
This question, in its most difficult form, is what we address in this paper.
Clustering nodes within a single network, also known as community detection, has been studied extensively.
Clustering nodes within multiple (related) networks, so-called multilayer or multiplex community detection, has also been studied.
Much less has been done, however, on trying to find structure among networks at the same time as performing community detection on them.

There are two fundamental network settings in terms of difficulty:
\begin{enumerate}[1.]
    \item There is \emph{node correspondence} among the networks. That is, the networks describe the relationship among the ``same'' set of nodes and we know the 1-1 correspondences that map 
    the nodes from network to network.
    We refer to this case as \emph{labeled}.
    \item The \emph{unlabeled} case, where there is no node correspondence among the networks. This includes the case where the nodes are the same among networks, but we do not know the correspondence. It also includes the much more interesting case where the nodes in each network are genuinely different; in this case, the networks can even be of different orders.
\end{enumerate}

We are interested in the unlabeled case.
Here, even if one performs individual clustering on nodes in each network, the clustering of the networks is still challenging, since there is the \emph{hard problem of matching} the estimated communities among different networks.
This is in addition to the information loss incurred by doing individual community detection.

Can we do simultaneous community detection in each network, borrowing information across them in doing so, and at the same time cluster the networks into groups themselves, all in the unlabeled setting? The nested stochastic block model (NSBM)  we propose in this paper is an affirmative answer to this question.

NSBM is a hierarchical nonparametric Bayesian model, extending the very well-known stochastic block model (SBM) for individual community detection to the setting of simultaneous network and node clustering.
In NSBM, the individual networks follow an SBM, but the node label (community assignment) priors and the connectivity matrices of these SBMs are connected through a hierarchical model inspired by the nested Dirichlet Process (NDP) \citep{rodriguez2008nested}.

In extending NDP to the network setting, we encountered difficulties and interesting phenomena when Gibbs sampling.
These issues exist in the original NDP, but are exacerbated, hence easier to observe, once we extend to the network setting.
To investigate these issues, we propose four different Gibbs samplers and provide an extensive numerical comparison among them.
A clear empirical picture arises through our simulations and real-world experiments as to the relative standing of these approaches.
Although past work has noted difficulties in sampling from DP mixtures, those involved in sampling NDP-based models seem to be of different nature.
As far as we know, our work is the first to report on these issues and provide a comprehensive exploration.
We provide a summary of our findings in Section~\ref{sec:sim}, suggest some plausible explanations, and leave the door open to future exploration of these phenomena, from both theoretical and practical perspectives.

\subsection{Related work}\label{sec:related}

There is a large literature on community detection for \emph{a single network}.
Approaches include graph partitioning, hierarchical clustering, spectral clustering, semidefinite programming, likelihood and modularity-based approaches, and testing algorithms~\citep{abbe2017community, fortunato2010community}.
There are also several Bayesian models that use the Dirichlet process for community detection in a single network \citep{kemp, meta, zhou2015infinite, newman2016structure, pmlr-v70-zhao17a, shen2022bayesian}.

Recently, there has been an emerging line of work on multiple-network data in both supervised and unsupervised settings including averaging \citep{ginestet2017hypothesis}, hypothesis testing \citep{chen2019hypothesis, chen2020spectral}, classification \citep{relion2019network, josephs2023bayesian}, shared community detection~\citep{lei2020consistent}, supervised shared community detection~\citep{arroyo2020simultaneous}, and shared latent space modeling~\citep{arroyo2021inference}.
Each of these methods is for the labeled case, but unlabeled networks arise in many applications involving anonymized networks, misalignment from experimental mismeasurements, and active learning for sampling nodes.
More generally, they arise as a population of networks sampled from a general \emph{exchangeable random array} model, including but not limited to graphon models~\citep{orbanz2014bayesian}.
While there is a large literature on graph matching in which the alignment between two networks is unknown, there are only a few methods for multiple networks in the unlabeled case \citep{kolaczyk2020averages, josephs2021network, amini2022hierarchical}.

\begin{table}[t]
\begin{adjustbox}{max width=\textwidth}
\begin{tabular}{lP{2cm}P{2cm}P{2cm}P{2cm}P{2cm}}
\toprule
& \textbf{Network\newline clustering} & \textbf{Node\newline clustering} & \textbf{Unlabeled\newline networks} & \textbf{Community\newline heterogeneity} & \textbf{Learns}\newline $K$ \\ \midrule
NSBM (this paper)& $\checkmark$ & $\checkmark$ & $\checkmark$ & $\checkmark$ & $\checkmark$ \\ \midrule
\cite{reyes2016stochastic} & $\checkmark$ & $\checkmark$ & & $(\checkmark)$ & $\checkmark$ \\
\midrule
sMLSBM \citep{stanley2016clustering} & \multirow{4}{*}{$\checkmark$} & \multirow{4}{*}{$\checkmark$} & & \multirow{4}{*}{$(\checkmark)$} & \\
MMLSBM \citep{jing2021community} & & & & & \\
ALMA~\citep{fan2022alma} & & & & & \\
\cite{signorelli2020model} &  &  & & &  \\
\midrule
NCLM \citep{mukherjee2017clustering} & $\checkmark$ & & $\checkmark$ & &  \\ \midrule
NCGE \citep{mukherjee2017clustering} & $\checkmark$ & &  &  &  \\ \midrule
\citet{diquigiovanni2019analysis} & $\checkmark$  & &  &  & \\ \midrule
RESBM \citep{paul2020random} & & $\checkmark$ & & $\checkmark$ &  \\ \midrule
\cite{mantziou2021bayesian} & $\checkmark$ & &  & $(\checkmark)$ & \\
\midrule
\cite{young2022clustering} & $\checkmark$ & &  & & \\ \midrule
HSBM \citep{amini2022hierarchical} & & $\checkmark$ & $\checkmark$ & $\checkmark$ & $\checkmark$ \\
\bottomrule
\end{tabular}
\end{adjustbox}

\medskip
\caption{Comparison of network clustering methods. $(\checkmark)$ in the penultimate column means the method allows community heterogeneity only at the level of network clusters.}
\label{tab:features}
\end{table}

There are several methods within the multiple-network inference literature for clustering a population of networks.
\citet{reyes2016stochastic} introduce a Bayesian nonparametric model for clustering networks with similar community structure over a fixed vertex set.
The authors modify the infinite relational model (IRM) from \citet{kemp}, which is closely related to the Dirichlet process, in order to capture (dis)assortative mixing and to introduce transitivity.

Motivated by soccer team playing styles, where each team is represented by a network, \citet{diquigiovanni2019analysis} introduce agglomerative hierarchical clustering of networks based on the similarity of their community structures, as measured by the Rand index.
The community structures are obtained by the  Louvain method.
The approach clearly requires a shared vertex set.

\citet{mukherjee2017clustering} provide two graph clustering algorithms: Network Clustering based on
\begin{enumerate*}[(a)]
    \item  Graphon Estimation (NCGE), and
    \item  Log Moments (NCLM).
\end{enumerate*}
Both methods perform spectral clustering on a matrix of pairwise distances among the networks.
In NCGE, one estimates a graphon for each network using, for example, neighborhood smoothing \citep{zhang_et_al} or universal singular value thresholding \citep{usvt} -- more accurately, they estimate $P = \ex[A]$ where $A$ is the $n\times n$ adjacency matrix.
The pairwise distances are the Frobenius distances of the estimated graphons.
In NCLM, one computes a vector of log moments for each network, $\bigl(\log \tr((A/n)^k), k=1,\dots,K\bigr)$, and then forms pairwise distances among these feature vectors.
NCGE only works in the labeled case, while NCLM can handle unlabeled case.

To further classify this literature, let us refine our earlier classification of network problems into the ``labeled'' and ``unlabeled'' cases.
For a community detection problem, the labeled case can be refined based on whether communities of the same (labeled) nodes are allowed to change across networks.
If a method allows such variation, we say that it can handle \emph{community heterogeneity}.
An example is the Random Effect SBM (RE-SBM) of~\citet{paul2020random}, where the communities across networks are Markov perturbations of a representative mean; we consider a very similar model in our simulations~ (Section~\ref{sec:sim}). Their work also allows variation in the off-diagonal entries of the SBM connectivity matrices. A very similar setting is considered in~\citet{chen2022global}, where minimax rates for recovering the global and individualized communities are established.

An alternative approach to modeling a population of networks is to assume that they are perturbations of some latent (true) network.
Letting $A^*$ denote the adjacency matrix of the true network, the simplest way to generate perturbed networks is to flip each entry $A^*_{ij}$ independently with some probability.
Among the first to consider this model are \citet{pedarsani2011privacy} who propose a simple version for studying the privacy of anonymized networks.
More recently, \citet{le2018estimating} generate perturbed $A = (A_{ij})$ via
\begin{align}
    \begin{split}\label{eq:ME}
    A_{ij} \given  A^*_{ij} \;\sim\;
    \begin{cases}
    \bern(1-Q_{ij}) & A^*_{ij} = 1\\
     \bern(P_{ij}) & A^*_{ij} = 0
    \end{cases}.
    \end{split}
\end{align}
Let us refer to the above as the measurement error (ME) process.
\citet{le2018estimating} assume that there is single latent $A^*$ generated from an SBM and one observes  multiple noisy measurements of it from the ME process above.
The authors also assume that $P$ and $Q$ matrices share the same block structure with the underlying SBM, and propose and analyze an EM type recovery algorithm.
We refer to the model from \citet{le2018estimating} as ME-SBM. 
\citet{mantziou2021bayesian} extend this model to allow multiple true networks $A^*_c, c=1,\dots,C$, each observed network being a perturbed version of one of them.
In other words, they consider a mixture of ME-SBMs and devise an MCMC sampler for the posterior, similar to the scheme in~\citet{lunagomez2021modeling}.
A similar mixture of (simplified) ME processes is considered in~\citet{young2022clustering} for network clustering in which $1-Q_{ij} = \alpha$ and $P_{ij} = \beta$.
The underlying latent networks $A_c^*$ are not assumed to have specific structures (such as SBM) and are inferred by Gibbs sampling.
Though the above works all assume labeled networks, \citet{josephs2021network} recently extended the mixture of ME processes to the unlabeled setting.

To summarize the previous two general approaches, in the RE-SBM, the ``communities'' undergo Markov perturbations to account for the variation in the observed networks, while in the ME-SBM type models, the ``adjacency matrices'' undergo a Markov perturbation.

A third line of work treats the multiple networks as layers of a single multilayer network.
Two notable example are the strata multilayer stochastic block model (sMLSBM) of~\citet{stanley2016clustering} and the  mixture multilayer stochastic block model (MMLSBM) of~\citet{jing2021community}, which are essentially the same model, independently proposed in different communities.
In both cases, the multiple networks, $A^j, j = 1,\dots,J$ -- viewed as layers -- are assumed to be independently drawn  from a mixture of SBMs, that is,
\begin{align}\label{eq:sbm:mix}
    A^j \sim \sum_{k=1}^K \pi_k \sbm(B_k, \bm \zeta_k) \enskip ,
\end{align}
where $\sbm(B_k, \bm\zeta_k)$ represent the distribution of the adjacency matrix of an SBM with connectivity matrix $B_k$ and label vector $\bm \zeta_k$.
This model allows community heterogeneity, but only at the level of network clusters, not at the level of individual networks.
To fit the model, \citet{stanley2016clustering} use a variational EM approach, while \citet{jing2021community} use tensor factorization ideas -- they compute an approximate Tucker decomposition via alternating power iterations, which they call TWIST -- and provide a theoretical analysis of their approach.  \citet{fan2022alma} introduce a slightly different ``alternating minimization algorithm (ALMA)'' to compute the tensor decomposition for MMLSBM, arguing that ALMA has higher accuracy numerically and theoretically compared to TWIST.
We compare with ALMA in Section~\ref{sec:sim}.
While MMLSBM can be used to simultaneously cluster networks and nodes, it requires the networks to be on a shared vertex set.
Along the same lines, \citet{signorelli2020model} model a population of networks as a general mixture model $A^j \sim \sum_{k=1}^K \pi_k f(\cdot \given\theta_k)$, but in the end focus mainly on the case where $f(\cdot\given \theta_k)$ represents either an SBM as in \eqref{eq:sbm:mix} or a $p_1$ model~\citep{holland1981exponential}.
They use EM for fitting the model and AIC/BIC to determine the number of components.

Besides the method of \citet{reyes2016stochastic}, all of the aforementioned methods assume the number of classes and communities to be known.
In contrast, by employing an NDP prior, our model learns the number of network classes and node communities.
Furthermore, as we will demonstrate later, our method does not require node correspondence between the observed networks and allows for community heterogeneity between networks even in the same class.
No other model in the literature exhibits these flexibilities.
A summary of features for various graph clustering methods is given in Table~\ref{tab:features} and more details on the advantages of NSBM are given in Section~\ref{sec:advantages}.

\subsection{Organization}

The remainder of our paper is organized as follows.
In Section~\ref{sec:ndp_and_nsbm}, we provide a review of the nested Dirichlet process before introducing NSBM.
We develop four efficient Gibbs algorithms for sampling from our posterior in Section~\ref{sec:Gibbs}, highlighting the nontrivial challenges from introducing dependency through the network data.
Section~\ref{sec:sim} is devoted to an extensive simulation study that compares our four samplers to competing methods on clustering problems of varying hardness.
We illustrate an important application of our model to two real datasets in Section~\ref{sec:data}.
Section~\ref{sec:conclusion} concludes our paper with a discussion of future work.

\section{Nested stochastic block model}\label{sec:ndp_and_nsbm}

\subsection{Original nested Dirichlet process}
\label{sec:original_ndp}

In a multicenter study, $y_{ij}$ is the observation on subject $i$ in center $j$.
For example, $\boldsymbol y_j$ is the vector of outcomes for patients in hospital $j$.
To analyze this data, it is common to either pool the subjects from the different centers or separately analyze the centers.
As a middle approach, \citet{rodriguez2008nested} introduce the nested Dirichlet process (NDP) mixture model for borrowing information across centers while also clustering similar centers.
That is, if $z_j$ is the hospital type for hospital $j$ and $\xi_{ij}$ is the patient type for patient $i$, then the NDP mixture model allows simultaneous inference on $\zb = (z_1, \ldots, z_J)$ and $\xib_j = (\xi_{1j}, \ldots, \xi_{n_jj}), \ j \in [J]$.

The original NDP mixture model can be expressed as follows\footnote{Here, we correct a minor, but subtle, error in the original paper. There, $Q$ is stated to be $\equiv \DP(\alpha \DP(\beta H))$, whereas one has to sample from this nested DP to get $Q$.}
\begin{align}
	Q &\sim \DP(\alpha \DP(\beta H)), \notag \\
	G_j \given Q &\sim Q, \notag \\
	y_{ij} \given G_j &\sim \int p(\,\cdot \given \theta) \, G_j(d\theta) \enskip ,
 \label{eq:ndp:y:mix}
\end{align}
where $j = 1,\dots,J$ in the second line and $i=1,\dots,n_j$ in the third line.
It is not immediately clear how this abstract version of the model can be extended to networks.
Below, we develop an alternative equivalent representation that is suitable for such extension.

Using the stick-breaking representation of the DP, we can explicitly write $Q$ as
\begin{align*}
	Q = \sum_{j=1}^\infty \pi_k \delta_{G_k^*}, \quad G_k^* \sim \DP(\beta H), \quad \pib \sim  \gem(\alpha) \enskip ,
\end{align*}
where $\pib = (\pi_k, \ k \in \nats)$.
Here, GEM stands for Griffiths, Engen, and McCloskey~\citep{pitman2002combinatorial} and refers to the distribution of a random measure on $\mathbb N$ stemming from the well-known stick-breaking construction of the DP~\citep{sethuraman1994constructive}.
See Section~\ref{sec:nsbm} for a more explicit description of GEM.
Another application of the stick-breaking representation, this time for $G^*_k$, gives us
\begin{align*}
	G^*_k =  \sum_{\ell=1}^\infty w_{\ell k} \delta_{\theta^*_{\ell k}}, \quad \theta^*_{\ell k} \sim H, \quad \wb_k \sim \gem(\beta) \enskip ,
\end{align*}
where $\wb_k = (w_{\ell k}, \ \ell \in \nats)$.

Next, we note \eqref{eq:ndp:y:mix} can be made more explicit by sampling $\theta_{ij} \given G_j \sim G_j$ i.i.d. over $i$, and then sampling $y_{ij}  \given \theta_{ij} \sim p(\,\cdot \mid \theta_{ij})$.
Furthermore, let $z_j$ denote the ``class'' of the $j$th center, i.e., which component of $Q$ gets assigned to $G_j$.
More specifically, $z_j = k$ iff $G_j = G^*_k$.

With this notation, $\theta_{ij} = \theta^*_{\xi_{ij}, z_j}$ and $G_j = G^*_{z_j}$, and the model reduces to
\begin{align}
    \theta^*_{\ell k} &\sim H
    & 
    y_{ij} \given \thb^*, z_k, \xi_{ij} &\sim p(\,\cdot \mid\theta^*_{\xi_{ij}, z_j}). \\
    \wb_k &\sim \gem(\beta) 
    &
    \xi_{ij} \given \wb, z_j &\sim \wb_{z_j} \\
    \pib &\sim \gem(\alpha) 
    &
    z_j \given \pib &\sim \pib 
\end{align}
We have tried to stay close to the notation in \citet{rodriguez2008nested}, with minor modifications (including changing $\zeta_j$ to $z_j$). For future developments, we will rename $\alpha$ to $\pi_0$ and $\beta$ to $w_0$.

\subsection{Degeneracy and non-identifiability of NDP}

Before we proceed with introducing NSBM, a few comments are in order.

\paragraph{Degeneracy} First, although our model mirrors the standard NDP mixture model as specified in \eqref{eq:ndp:y:mix}, there is an important difference related to the notion of ``shared atoms'' in the NDP. The original NDP was designed for i.i.d. data, with the apparent motivation of allowing for sharing atoms (patient types) among mixing measures ($G^*_k$) of different high-level clusters (hospital types). For instance, we might want two different hospital types $G^*_1$ and $G^*_2$ to share some but not all patient types, e.g. $G^*_1 = \frac12 \delta_{\theta^*_1} + \frac12 \delta_{\theta^*_2}$ and $G^*_2 = \frac12 \delta_{\theta^*_1} + \frac12 \delta_{\theta^*_3}$ where $\theta^*_1, \theta^*_2$, and $\theta^*_3$ are different. As recently noted by \citet{camerlenghi2019latent}, this is not possible in the NDP if the underlying measure $H$ is non-atomic. In the non-atomic case, the prior does not allow two different hospital types $G^*_1$ and $G^*_2$ to share an atom since $\theta^*_{\ell k}$ for $\ell, k \in \nats$ are i.i.d. draws from $H$ and hence all distinct with probability 1. This degeneracy in the prior leads to the following empirical posterior behavior: Either the model recognizes the common atom $\theta^*_1$, collapsing $G^*_1$ and $G^*_2$ into a single cluster (with atoms $\theta^*_1$, $\theta^*_2$, and $\theta^*_3$), or it correctly identifies two distinct bimodal clusters but with distinct atoms. While this behavior is consistent with NDP, it may not align with its original design intentions.  It is perhaps best understood as a case of model misspecification.

In contrast to the i.i.d. setting, in our network setting, sharing atoms among high-level clusters is not viable, making this degeneracy irrelevant. This is because we model networks as SBMs, where an atom corresponds to a row of the SBM connectivity matrix -- a pattern of how a given community connects to other communities. Two different SBMs can have identical rows without implying the communities are the same, especially when other rows differ.

Consider, for example, the following two different SBMs: 
\begin{equation}\label{eq:shared_atoms} 
    \etab_1 = 
    \begin{pmatrix} 
    p & q \\ q & r_1 
    \end{pmatrix} \quad~\text{and}~\quad 
    \etab_2 = 
    \begin{pmatrix} 
    p & q \\ q & r_2 
    \end{pmatrix}, 
\end{equation} where $r_1 \neq r_2$. While one might be tempted to declare the first community in $\etab_1$ and $\etab_2$ identical given their matching connection patterns $(p, q)$, this interpretation is not natural for networks. Communities in different SBMs do not necessarily carry the same meaning, and matching values are more coincidental than meaningful. Moreover, even if we consider community 1 the same in both SBMs, the atom $(p, q)$ differs between them since community 2 differs (implied by $r_1 \neq r_2$), meaning $q$ measures connection strength to distinct entities.

Following this logic, two SBMs that differ in even a single row/column should be considered as having no shared atoms, which is exactly what NDP models. What is termed NDP degeneracy is precisely what we want in the network setting: networks are clustered together only if they come from the same underlying SBM, regardless of partial entry matches with other SBMs.

\paragraph{Non-identifiability}Second, there is an inherent non-identifiability in NDP mixture models that has not been previously discussed in the literature.
In the original NDP motivating example, this amounts to not being able to differentiate between a multicenter study with $K$ hospital types and $L_k$ patient types in the $k$th hospital type versus a single hospital with $L = \sum_{k=1}^K L_k$ patient types.
In other words, since a single hospital can have infinitely many patient types, the likelihood of the NDP cannot distinguish these two cases. One has to rely solely on the strength of the prior to differentiate them.
While this issue is less apparent in the original NDP for Euclidean data, we will see that introducing network data exposes this challenge.
Again, this is distinct from degeneracy; Degeneracy is an issue with the NDP \textit{prior}, whereas non-identifiability is about the \textit{likelihood}.

\subsection{NSBM} \label{sec:nsbm}

Here, we introduce the nested stochastic block model (NSBM), which is a novel employment of the NDP as a prior for modeling the two-level clustering structure on a collection of network objects.

Let $\boldsymbol A = A^1, \ldots, A^J$ be the observed adjacency matrices of $J$ networks (say from $J$ subjects) in which $A^j$ has $n_j$ nodes.
Our goal is to model both the within and between clustering structures of the networks.
That is, we want to group a collection of network objects into classes while simultaneously clustering the nodes within each network into communities.

Let $K$ be the number of classes which is not known or specified.
We denote $\boldsymbol z~=~(z_1, \ldots, z_J)$ with $z_j \in \{1,\ldots, K\}$ as the class membership for the $j$th network.
Given the partition structure $\boldsymbol z$ of these objects, we assume that each of the networks in each class of networks follows a stochastic block model (SBM) with $n_j$ nodes.
Denote by $\boldsymbol \xi_j=(\xi_{1j}, \ldots, \xi_{n_jj})$ the community membership of the $n_j$ nodes for the $j$th network; $\xib_j$ encodes the clustering structure of the nodes within the $j$th network.

We assume that we can borrow information and cluster across distributions by imposing that $\zb$ and 
$\xib_j, \ j \in [J]$ follow an NDP prior.
This leads to the following \emph{nested SBM}:
\begin{align}
    \eta_{x yk} &\sim \Beta(\alpha, \beta) 
    & 
    A^j_{st} \given \xib_j, z_j, \etab &\sim \bern(\eta_{\xi_{sj} \xi_{tj} z_j}) \label{eq:sbm} \\
    \wb_k &\sim \gem(w_0)  \label{eq:downlevel}
    &
    \xi_{sj} \given \wb, z_j &\sim \wb_{z_j} \\
    \pib &\sim \gem(\pi_0) 
    &
    z_j \given \pib &\sim \pib \label{eq:uplevel}
\end{align}
independently over $1\le s < t \le n_j$, $x \le y$, $k \in \nats$ and $j \in [J]$. Both $x,y$ range over $\nats$ subject to $x \le y$. 
We note that~\eqref{eq:sbm} specifies the SBM likelihood, \eqref{eq:downlevel} models the community structure within each network, and \eqref{eq:uplevel} models the class level of the network objects.
If $\sbm(B, \bm\zeta)$ represent the distribution of the adjacency matrix of an SBM with connectivity matrix $B$ and label vector $\bm \zeta$, we can compactly write the RHS of \eqref{eq:sbm} as 
\[
A^j \given \xib_j, z_j, \etab \sim \sbm(\etab_{z_j}, \xib_j) \enskip .
\]

The sampling of $\pib$ and $\wb_k$ in lines~\eqref{eq:downlevel}--\eqref{eq:uplevel} can be made more explicit as follows:
\begin{alignat}{2}
	u_{x k} &\sim \Beta(1,w_0), &\quad \wb_k &= F(\ub_k), \\
	v_{k} &\sim \Beta(1,\pi_0), &\quad \pib &= F(\vb) \enskip ,
\end{alignat}
independently across $x, k \in \nats$,
where $\ub_k = (u_{xk})$, and $\vb = (v_k)$, and $F(\cdot)$ is the stick-breaking map.
More specifically, $F : [0,1]^\nats \to [0,1]^\nats$ is given by
\begin{align}
    [F(\vb)]_k := v_k \prod_{\ell=1}^{k-1} (1-v_\ell) \enskip ,
\end{align}
where, by convention, the product over an empty set is equal to 1.

We will use the following conventions for indices.
We often use $s,t$ to index nodes, and use $x, y$ to index communities within networks. We use $k$ to index the class of networks themselves.
Throughout, we maintain the notation that $j$ is the layer/network index, $z_j$ is the class of network $j$,
$\xi_{sj}$ is the community label for the $s^{th}$ node in the $j^{th}$ network, and $\etab_k$ is the connectivity matrix for the $k$th SBM.

\subsection{Advantages}\label{sec:advantages}

Here, we remark on some of the advantages of NSBM.
First, NSBM is the only existing method that currently achieves within and between network clustering \emph{simultaneously} for networks.
By employing an NDP prior, our model allows one to borrow strength from networks that are in the same class in performing clustering.
This is in contrast to two-step procedures (such as NCGE and NCLM) that first cluster the network objects into classes and then, assuming the networks share a common vertex set, cluster the nodes of networks in a given class into communities.

Moreover, unlike these two-step procedures, our method assigns communities to the nodes for each network within a given class individually.
That is, the communities are inferred on a network level rather than a class level, which we refer to as \textit{community heterogeneity}.
Community heterogeneity can occur either when nodes change communities between networks, for example if an individual changes friend groups, or if the distribution of nodes in each community differs across networks.
In both of these cases, the heterogeneity exists \textit{within} the same network class.

Allowing for community heterogeneity can be easily seen from our construction.
Specifically, the community vectors $\xib_j = (\xib_{tj})_{t=1}^{n_j}$ are inferred at the node level separately for each network.
This also underscores the feature that the collection or sample of networks does not have to share a common set of nodes.
In case the vertex set is the same, or even if just the number of nodes is the same, our construction does not require a correspondence between the node labels.
In the literature, these networks are considered \textit{unlabeled}.

While the distinction between labeled and unlabeled networks has received the most attention in the multiple network literature, community heterogeneity is a distinct but important feature.
Ultimately, these features make our setup much more general and flexible compared to any of the other existing methods.
These features are necessary in the common scenario, for example with social networks, in which the networks lack a node correspondence (anonymity) and/or have a different number of actors (different sample).

Finally, our model does not require the prespecification of the number of classes or communities.
The value of this feature cannot be overstated, since methods that require $K$ to be given as input necessarily rely on heuristics such as eigenvalue gaps or elbow plots that are not theoretically justified.

\paragraph{A note on sparsity}  Our model naturally accommodates sparse networks through the edge probability parameters $\eta_{xyk}$. While network sizes $n_j$ influence the total number of potential connections, the model captures varying levels of sparsity by adjusting these probability parameters downward. The current formulation assumes that networks of the same type share similar sparsity characteristics, which is reasonable in many applications where connection probabilities are determined by node types and network class. For applications requiring greater sparsity-heterogeneity among networks of the same type, the model can be extended by introducing network-specific scaling parameters $\gamma_j$ that multiply existing edge probabilities $\eta_{\xi_{sj} \xi_{tj} z_j}$, where $\gamma_j$ would receive a Beta prior to constrain values to $[0,1]$.

\subsection{Joint distribution}\label{sec:joint}

Recall that we observe multiple networks with adjacency matrices  $A^j, \ j \in [J]$, from SBMs with connectivity matrices $\etab_k~=~(\eta_{xyk}), \ k \in~\nats$.
The joint density of NSBM factorizes as
\begin{multline}\label{eq:joint}
    p(\boldsymbol A, \etab, \xib, \zb, \ub, \vb)
    = \prod_j \Big[ \pi_{z_j}
    \prod_{s < t} p(A^j_{st} \given \etab, \xib_j, z_j) 
    \prod_s p(\xi_{sj} \given \wb, z_j) 
    \Big] \cdot \\ 
      \Bigl[   
    p(v_k) 
    \prod_{x} p(u_{xk}) 
    \prod_{x\le y} p(\eta_{xyk})    
    \Bigr]
\end{multline}
where $p(\xi_{sj} \given \wb, z_j) = w_{\xi_{sj}, z_j}$ and $p(A^j_{st} \given \etab, \xib_j, z_j)= 
\eta_{\xi_{sj} \xi_{tj} z_j}^{A_{st}^j}(1-\eta_{\xi_{sj} \xi_{tj} z_j})^{1-A_{st}^j}$. 
\begin{align}\label{eq:Gamma:index:3}
	\Gamma_{xy}^j = 
	\begin{cases}
		\{(s,t) : 1 \le s < t \le n_j\}, &x = y \\
		\{(s,t) : 1 \le s \neq t \le n_j\} & x \neq y
	\end{cases} \enskip ,
\end{align}
and define the block sums
\begin{align*}
	\m_{xy}^j &= \sum_{(s,t) \in \Gamma_{x y}^j} A_{st}^j \ind{\xi_{sj} = x, \xi_{tj} = y}, &
	N_{xy}^j &= \sum_{(s,t) \in \Gamma_{x y}^j} \ind{\xi_{sj} = x, \xi_{tj} = y},
\end{align*}
and	the \emph{aggregate} block sums
\begin{align*}
	\m_{xyk} &= \sum_j \ind{z_j = k} m_{xy}^j, &
	N_{xyk} &= \sum_j  \ind{z_j = k} N_{xy}^j, &
        \bar m_{xyk} &= N_{xyk} - m_{xyk}.
\end{align*}
Then 
\begin{align}
	\prod_j \prod_{s <t} p(A_{st}^j \given \etab, \xib_{j},z_j ) &= 
	\prod_j \prod_{x \le y} \eta_{xy z_j}^{m_{xy}^j} (1-\eta_{xy z_j})^{N_{xy}^j - m_{xy}^j} \label{eq:j:product}\\
	&=\prod_k \prod_{x \le y} \eta_{xyk}^{m_{xyk}} (1-\eta_{xyk})^{N_{xyk} - m_{xyk}}\label{eq:k:product}
\end{align}
and
\begin{align}
	p(\xi_{sj} \given \cdots) &\;\propto\; p(\xi_{sj} \given \wb, z_j) \prod_{t: t\neq s} p(A_{st}^j \given \etab, \xib_j, z_j) \notag \\
	&\;\propto\; w_{\xi_{sj}, z_j} \prod_y \prod_{t: t\neq s, \, \xi_{tj} = y} 
	\eta_{\xi_{sj} y z_j}^{A_{st}^j} (1- \eta_{\xi_{sj} y z_j})^{1-A_{st}^j} \label{eq:xi:s:j:cond}.
\end{align}

These compact representations allow efficient sampling, which we describe in the next section.

\section{Posterior inference}\label{sec:Gibbs}

To sample from the posterior distribution, we propose the use of samplers based on truncation approximations, which is what was used in the original Gibbs sampler for NDP.
To do so, we suppose there is a finite number of classes $K$ and clusters of nodes within class, $L$, with $K$ and $L$ specified as very large numbers.

There are many known challenges in the literature on sampling from DP mixture models, in particular with the stick-breaking representation, including mixing issues due to local modes and sensitivity to initialization \citep{neal2000markov, hastie2015sampling}.
As mentioned earlier, NSBM introduces new challenges due to the non-identifiability inherent in the NDP and the complexity of the SBM likelihood.
To investigate these issues, we consider and compare the following four samplers:

\begin{enumerate}[(a), itemsep=1ex]
    \item \textbf{Gibbs (G)}: A standard Gibbs samplers that updates all the (vector) parameters $\xib, \zb, \etab, \wb$ and $\pib$, one at a time given all the rest.
    \item \textbf{Collapsed Gibbs (CG)}: Similar to (G), but we first marginalize out $\etab$ and only perform Gibbs updates on the remaining parameters.
    \item \textbf{Blocked Gibbs (BG)}: Same as (G) except that we sample $(\xib,\zb)$ jointly given the rest of the variables. This joint sampling is exact. Letting $E = (\etab, \wb, \pib)$ be all the parameters except the labels, we perform the joint sampling exactly, by sampling $\xib \given E$ first, followed by $\zb \given \xib, E$.
    \item \textbf{Incompatible Blocked Gibbs (IBG)}: Same as (BG) but with the order of the label updates reversed: First, we sample  $\zb \given \xib, E$ and then $\xib \given E$.
\end{enumerate}

For each of our samplers, we initiate $\xib$ using separate DPSBM models \citep{shen2022bayesian} on each network.

\subsection{Gibbs (G)}

Given the joint distribution in~\eqref{eq:joint}, we can easily compute the conditional distributions for a standard Gibbs sampler (G).

\begin{enumerate}[wide, labelindent=10pt, itemsep=1ex]
    \item \textbf{Sampling $\etab$} \quad Sample $\eta_{xyk}$ independently for $(x,y): x\le y$:	
	\begin{align*}
		\eta_{xyk} \given \cdots \;\sim\; \Beta\Bigl(m_{xyk} +\alpha, \bar m_{xyk} + \beta \Bigr)
	\end{align*}
	This follows from representation~\eqref{eq:k:product}.
 
	\item \textbf{Sampling $\xib$} \quad Sample $\xi_{sj}$ sequentially, given $\xi_{-sj}$ and other parameters, from the discrete distribution:
	\begin{align*}
		p(\xi_{sj} = x \given \cdots) &\propto w_{xz_j}
		\exp \Bigl( \sum_y
		\tau_{sy}^j u_{xyz_j} + \n_{sy}^j v_{xyz_j} \Bigr)
	\end{align*}
	where $u_{xyk} = \log[\eta_{xyk}/ (1-\eta_{xyk})]$ and $v_{xyk} = \log (1- \eta_{xyk})$ and
	\begin{align*}
		\tau_{sy}^j = \sum_{t:t \neq s} A_{st}^j  \ind{\xi_{tj} = y}, \quad 
		\n_{sy}^j =  \sum_{t:t \neq s} \ind{\xi_{tj} = y}.
	\end{align*}
	This follows from~\eqref{eq:xi:s:j:cond}.
	
	\item \textbf{Sampling $\zb$}\quad Sample $z_j$ independently (in parallel) over $j$ given everything else from
	\begin{align*}
		p(z_j  \given \cdots) &\propto 
		\pi_{z_j} 
		\prod_x w_{x z_j}^{n_x(\xi_j)} 
		\prod_{x \le y} \eta_{xy z_j}^{m_{xy}^j} (1-\eta_{xy z_j})^{N_{xy}^j - m_{xy}^j},
	\end{align*}
	that is,
	\begin{align*}
		p(z_j = r \given \cdots) &\propto  \pi_r \exp \Bigl( \sum_x n_x(\xib_j) \log w_{x r} +  \sum_{x \le y} 
		\bigl( m_{xy}^ju_{rxy} + N_{xy}^j v_{rxy}\bigr) \Bigr).
	\end{align*}
	where
	$
		n_x(\xib_j) = \{s: \; \xi_{sj} = x\}.
	$
	This follows from representation~\eqref{eq:j:product} and 
	\[
		\prod_s p(\xi_{sj} \given \wb, z_j) = 
		\prod_s w_{\xi_{sj}, z_j} = 
		\prod_x w_{xz_j}^{n_{x}(\xib_j)}.
	\]

	\item \textbf{Sampling $\ub$} \quad Sample $u_{xk}$ independently across $x$ and $k$, as
	\begin{align}\label{eq:u:update}
		u_{xk} \mid \cdots \sim 
		\Beta\bigl(n_{x}(\xib^{(k)}) + 1,  n_{>x}(\xib^{(k)}) + w_0\bigr),
	\end{align}
	where $\xib^{(k)} := \{\xi_{sj}: s \in [n_j],\; \,z_j = k\}$ and $n_x(\cdot)$ and $n_{ > x}(\cdot)$ are operators counting how many labels are equal to or greater than $x$, respectively. 
	This follows by noting that 
	\begin{align*}
		p(\ub \mid \cdots) &\; \propto \; 
		\prod_{s,j} w_{\xi_{sj}, z_j} \prod_{k,x} b_{w_0}(u_{x k})  \\
		&\;=\; \prod_k \Bigl[ \prod_{(s,j): z_j = k}  [F(\ub_k)]_{\xi_{sj}} \prod_x b_{w_0}(u_{x k}) \Bigr],
	\end{align*}
	where the second line uses
	$w_{\xi_{sj}, z_j} = [w_{z_j}]_{\xi_{sj}} = [F(\ub_{z_j})]_{\xi_{sj}}$.
    This shows that the posterior factorizes over $k$.
	
	\item \textbf{Sampling $\vb$} \quad Sample $v_k$ independently from 
	\begin{align}\label{eq:v:update}
			v_{k} \mid \cdots  \sim \Beta(n_{k}\big(\zb) + 1,  n_{>k}(\zb) + \pi_0\big).
	\end{align}
	This follows by noting that 
	\begin{align*}
		p(\vb \given \cdots) \propto \prod_j \pi_{z_j} \prod_k b_{\pi_0}(v_k) = \prod_j [F(\vb)]_{z_j} \prod_{k} b_{\pi_0}(v_k)
	\end{align*}
	and using the standard lemma.

 \end{enumerate}

\subsection{Collapsed Gibbs (CG)}

Since our goal is simultaneously clustering networks and nodes, the only parameters we need to infer are $\zb$ and $\xib$.
Hence, the underlying block matrices $\etab$ are nuisance parameters.
Following the Beta-Binomial conjugacy in our model, we can collapse out the link probabilities of each of the networks for more efficient updating.
Note that we retain the ability to estimate these probabilities as the proportion of observed edges between inferred communities in each class.

Recall that using the aggregate block sums in Section~\ref{sec:joint}, \eqref{eq:k:product} allowed us to factor the likelihood as a product over $k$.
Combining this with the prior on $\etab$, we have
\begin{align}\label{eq:eta:products}
	\propto \prod_k \prod_{x \le y} \eta_{xyk}^{m_{xyk} + \alpha - 1} \bigl(1-\eta_{xyk} \bigr)^{\bar m_{xyk} + \beta-1}.
\end{align} 
Integrating over $\etab$, we obtain the collapsed joint distribution which is given by
\begin{align}\label{eq:full:joint}
	p(\Ab,   \zb, \xib \mid  \ub, \vb) &\;\stackrel{\approx}{\propto}\; 
	\prod_k \prod_{x \le y} B\bigl(m_{xyk} + \alpha ,\; \bar m_{xyk} + \beta  \bigr)
	\prod_j \Big[ \pi_{z_j} \prod_{s=1}^{n_j} w_{\xi_{sj}, z_j}\Big],  \\  
	p(\ub, \vb) &\;\stackrel{\approx}{\propto}\; \prod_{k= 1}^K \Bigl[ b_{\pi_0}(v_k)  \prod_{x = 1}^L b_{w_0}(u_{x k}) \Big],
\end{align}
where $B(\cdot, \cdot)$ is the beta function.

This allows us to define a collapsed Gibbs sampler (CG), which is the same as (G) except for the following updates:

\begin{enumerate}[wide,labelwidth=!, labelindent=10pt, itemsep=1ex]
    \item[(2)] \textbf{Sampling $\xib$} \quad
    Given $\zb$, only $m_{xyz_j}$ and $\bar m_{xyz_j}$ depend on $\xi_{sj}$.
    Hence,
    \begin{align*}
	   p(\xi_{sj} \mid \cdots) \;\propto\;  w_{\xi_{sj},z_j}\prod_{x \le y} B\bigl(m_{xy}^{z_j} + \alpha_\eta ,\; \bar m_{xy}^{z_j} + \beta_\eta  \bigr).
    \end{align*}
    The complexity of computing the product above can be reduced from $O(L^2)$ to $O(L)$ by a beta ratio idea similar to the one described below for sampling $z_j$.

    \item[(3)] \textbf{Sampling $\zb$} \quad Let us introduce some notation to simplify the updates of $z_j$ given everything else.
    Assume that the previous value of $z_j = r_0$  and the new candidate value is $r$.
    Let $q_{xyk}$ and $\bar q_{xyk}$ denote the previous values of $m_{xyk}$ and $\bar m_{xyk}$ based on $z_j = r_0$, and let $m_{xyk}(r)$ and $\bar m_{xyk}(r)$ denote the new values based on $z_j = r$, where we are showing the internal dependence on $r$ explicitly. 

    Changing $z_j$ from $r_0$ to $r$ only changes two of the matrices $\{m_{xyk}(r), k~\in~[K]\}$, namely $m_{xyr}(r)$ and $m_{xyr_0}(r)$.
    When $r \neq r_0$, the effect is to subtract values from $m_{xyr_0}(r)$ and add them to $m_{xyr}(r)$ such that $m_{xyr_0}(r) + m_{xyr}(r)$ remains the same.
    The change to $m_{xyr}$ is the same for all $r \neq r_0$ and is equal to the block sums of $A^j$ with respect to $\xib_j$.
    For $ r = r_0$, of course the is no change to $m_{xyk}(r)$.
    
    More specifically, fix $j$ and let
    \begin{equation*}
        D_{xy}^j := \sum_{(s,t) \in \Gamma_{xy}^j} A_{st}^j \, \ind{\xi_{sj} = x,\, \xi_{tj} = y} \enskip .
    \end{equation*}
    Then, for all $r \neq r_0$,
    \begin{align}\label{eq:m:incremental:update}
	   \begin{cases}
	       m_{xyk}(r) = q_{xyk}, & k \not \in \{r,r_0\},\\
	       m_{xyr}(r) = q_{xyr} + D_{xy}^j, \\
	       m_{xyr_0}(r) = q_{xyr_0} - D_{xy}^j,
	   \end{cases}
    \end{align}
    and $m_{xyk}(r_0) = q_{xyk}$ for all $k$.
    Similar relations holds between $\bar m_{xyk}(r)$ and $\bar q_{xyk}$ with $\bar D_{xy}^j$ replacing $D_{xy}^j$ and defined by replacing $A_{st}^j$ with $\bar A_{st}^j$ in the definition of $D_{xy}^j$.

    Dividing~\eqref{eq:full:joint} by $\prod_k \prod_{x \le y} B\bigl(q_{xyk} + \alpha ,\; \bar q_{xyk} + \beta \bigr)$, which is constant when considering $p(z_j = r \mid \cdots)$, we obtain
    \begin{align}\label{eq:z:update}
    	\begin{split}
	   p(z_j = r \mid \cdots) \;\propto\; \pi_{r} \prod_{s=1}^{n_j} w_{\xi_{sj}, r} \prod_{k \in \{r_0, \, r\}}
		\prod_{x \le y} 
		\frac{ B\bigl(m_{xyk}(r) + \alpha ,\; \bar m_{xyk}(r) + \beta  \bigr) }
		{B\bigl(q_{xyk} + \alpha ,\; \bar q_{xyk} + \beta \bigr)}.
		\end{split}
    \end{align}
    where we have used~\eqref{eq:m:incremental:update} to simplify the product over $k$.

    Note from~\eqref{eq:m:incremental:update} that $m_{xyr_0}(r)$ is the same for all $r \neq r_0$, obtained by subtracting counts from $q_{xyr_0}$, while we have $m_{xyr_0}(r_0) = q_{xyr_0}$.
    
    Let
    \begin{align*}
	   \kappa_r := \prod_{x \le y}
		\frac
		{B\bigl(m_{xyr_0}(r) + \alpha ,\; \bar m_{xyr_0}(r) + \beta \bigr)}
		{B\bigl(q_{xyr_0} + \alpha ,\; \bar q_{xyr_0} + \beta \bigr)}, \quad r \neq r_0
    \end{align*}
    and $\kappa_{r_0} := 1$.
    Note that $\kappa_r$ is the same for all $r \neq r_0$ and can be computed once for some $r \neq r_0$.
    Considering the two terms $k = r_0$ and $k = r$ in~~\eqref{eq:z:update} separately, we have
    \begin{align}\label{eq:z:update:final}
	   p(z_j = r \mid \cdots) & \;\propto\;
	   \pi_{r} \prod_x w_{x r}^{n_x(\xib_j)} \cdot \kappa_r  \prod_{x \le y} 
	   \frac
	       {B\bigl(m_{xyr}(r) + \alpha ,\; \bar m_{xyr}(r) + \beta \bigr)}
	       {B\bigl(q_{xyr} + \alpha ,\; \bar q_{xyr} + \beta \bigr)},
    \end{align}
    using the further simplification $\prod_{s=1}^{n_j} w_{\xi_{sj},r} = \prod_x w_{xr}^{n_x(\xib_j)}$.
    Here, $n_x(\cdot)$ counts how many labels are equal to $x$.
    %
    The beta ratios in~\eqref{eq:z:update:final} can be computed efficiently and accurately, which we describe in Appendix~\ref{sec:efficiency}.
\end{enumerate}

\subsection{Blocked Gibbs (BG)}

\citet{rodriguez2008nested} proposed a Gibbs sampler when they introduced the NDP.
Although not mentioned, they used a block sampler for sampling exactly from the joint distribution of $\zb$ and $\xib$. Specifically, letting $E =(\yb, \thetab, \wb, \pib)$ collect all the parameters except $\xib$ and $\zb$, the authors' approach exactly samples from  $p(\zb, \xib \given E)$ as one step in their Gibbs sampler. Note that because $p(\zb, \xib \given E) = \prod_{j} p(z_j, \xib_j \given E)$ factorizes over $j$, it is enough to focus on a single $j$.

\citet{rodriguez2008nested} suggested marginalizing out $\xi_j$ to sample from $p(z_j \mid E)$ as: \begin{align*}
	p(z_j \given E) &= \sum_{\xib_j} p(z_j, \xib_j \given E) \\
	&\propto \pi_{z_j}  \sum_{\xib_j} \prod_i \bigl\{ p(y_{ij} \given \theta, \xi_{ij}, z_j)
	w_{\xi_{ij}, z_j} \bigr\} \enskip ,
\end{align*}
and then sampling from $p(\xib_j \given z_j , E)$.
By interchanging the sum and product, we obtain
\begin{align*}
	p(z_j \given E)
	&\propto  \pi_{z_j}  \prod_i \sum_{\xi_{ij}} \bigl\{ p(y_{ij} \given \theta, \xi_{ij}, z_j)
	w_{\xi_{ij}, z_j} \bigr\} \enskip .
\end{align*}
Importantly, this exchange is justified because of the following identity:
\begin{align*}
    \sum_{x_1,\dots,x_n} \prod_{i=1}^n f_i(x_i) 
    &= \sum_{x_1,\dots,x_{n-1}} \Bigl[ f_1(x_1) f_2(x_2)\cdots f_{n-1}(x_{n-1}) \sum_{x_n}f_n(x_n) \Big] \\
    &=\sum_{x_1,\dots,x_{n-1}} \Bigl[ f_1(x_1) f_2(x_2)\cdots f_{n-1}(x_{n-1}) \Big] \cdot \sum_{x_n}f_n(x_n) 
    \\ 
    &= \prod_{i=1}^n \sum_{x_i} f_i(x_i),
\end{align*}
where $x_i \in [L]$ for all $i$.
The last equality follows by a recursive application of the argument.
In the NDP, $f$ is the likelihood function and $\xi_{ij}$ plays the role of $x_i$ here.
Note the summation on the LHS is over $L^n$ terms.
The overall complexity of calculating the LHS directly is $O(n L^n)$, while that of the RHS is $nL + n = O(nL)$.

This interchange of summation and product does not work for the NSBM because our likelihood function does not factor as a product.
This is directly a result of the dependency introduced by modeling network data.
Specifically, our likelihood is a function of \textit{edges}, which encode the information between \textit{pairs} of nodes, rather than individual data points.

Unfortunately, exact sampling from $p(z_j \mid E)$ is intractable without the interchange.
In other words, the complexity of exact sampling from $p(z_j \given E)$ is that of sampling from the posterior of an SBM. See Section~\ref{sec:conclusion} for further discussion.
However, we can recover a blocked sampler by swapping the order: first sample $\xib_j$ from $p(\xib_j \given E)$ by marginalizing out $z_j$ and then sampe $z_j$ from $p(z_j \given \xib_j, E)$.
This allows us to utilize the interchange for NSBM.

We obtain the blocked Gibbs sampler (BG) by replacing the $\xib$ update in (G) with the following:
\begin{enumerate}[wide, labelindent=10pt, itemsep=1ex]
    \item[(2)] \textbf{Sampling $\xib$} \quad 
    Sample $\xi_{sj}$ sequentially, given $\xi_{-sj}$ and the other parameters besides $\zb$, from the discrete distribution:
	\begin{align*}
		p(\xi_{sj} = x \given \cdots) &\propto \prod_{k} \sum_y \pi_k \cdot w_{xk} \cdot \exp \Bigl(\tau_{sy}^j u_{xyz_j} + \n_{sy}^j v_{xyz_j} \Bigr)
	\end{align*}
\end{enumerate}

A blocked sampler is a particular type of partially collapsed Gibbs (PCG) sampler.
\citet{van2008partially} demonstrate the importance of sampling order when implementing PCG samplers.
In particular, we need to sample first from $p(\xib_j \mid E)$ and then from $p(z_j \mid \xib_j, E)$ since sampling in the opposite order may result in a stationary distribution that differs from our target posterior distribution.
However, it is often ``only" dependency between parameters that is lost, which may not affect model performance such as clustering results.
We therefore consider a fourth sampler, which we refer to as an incompatible blocked Gibbs (IBG), that uses the opposite sampling order.

\subsection{Initialization}

In theory, each sampler will have a stationary distribution equal to our posterior regardless of how the parameters are initialized.
In practice, we find that the performance is improved with a ``warm-start" for the $\bm \xi$ labels, i.e. a non-random initialization.
In particular, we initialize the $\bm \xi$ labels by separately fitting a DP-SBM on every network.
This is a natural initialization for NSBM because it effectively assigns every network a separate class and then fits a DP-SBM at each layer.

We note that for random initializations, (CG) performs the same as its non-random counterpart, whereas all of the other samplers often get stuck at a single network class. This is because of the order of our updates. When we randomly initialize $\bm \xi$, the next step for all of the non-collapsed samplers is to update $\bm \eta$. If the $\bm \xi$ labels are wrong, which we expect with a random intialization, then $\bm \eta$ is going to be a bad estimate of the true underlying probability matrices. On the other hand, (CG) marginalizes out $\bm \eta$, which is why the random initialization still works.

\section{Simulations}\label{sec:sim}

In this section, we perform several simulation studies to evaluate the clustering performance of NSBM.
We first provide a simulation that analyzes the various Gibbs samplers from Section~\ref{sec:Gibbs}.
We then assess how each NSBM sampler scales as the number of nodes grows.
Lastly, we compare the performance of NSBM to three competing methods on both assortative and non-assortative networks.
The first two comparison methods are two-step procedures based on the graph clustering algorithms from \citet{mukherjee2017clustering}: first cluster the networks using NCGE or NCLM, then cluster the nodes using a community detection algorithm on the average adjacency matrix from the networks in each class.
Although NCLM works for unlabeled networks, the second step requires the networks to be labeled so that averaging the adjacency matrices is sensical.
We also compare our method to MMLSBM from \citet{jing2021community} using the improved algorithm ALMA from \citet{fan2022alma}.
While this method simultaneously clusters networks and nodes, it also requires the networks to be labeled.

We compare our model with these three methods in various simulation settings.
In such settings, we sample labeled networks so that the comparison to the other methods is fair.
However, we emphasize that NSBM does not utilize the node correspondence, hence, as we show in Section~\ref{sec:sim_heterogeneity}, the performance would be the same if the networks were not labeled.
For NCGE, NCLM, and ALMA, we provide the true number of classes and clusters as input.
In contrast, NSBM learns both the number of network classes and the number of node communities automatically.

For NSBM, we take flat hyperpriors for $\bm \eta, \bm w,$ and $\bm \pi$, i.e. $\eta_{xyk} \sim \text{Beta}(1, 1)$ and $w_0 = \pi_0 = 1$.
We then summarize the posterior using the labels that minimize the posterior expected Variation of Information (VI) using the \texttt{salso} package  \citep{dahl2022search}.
We measure the performance of the estimated $\zb$ and $\xib$ labels using normalized mutual information (NMI).
NMI ranges from 0 (random guessing) to 1 (perfect agreement) and is used to compare two partitions with a different number of labels while accounting for the issue of label invariance.
For $\xib$-NMI, we report the average NMI across $\xib_j$.

The simulations were performed on a high-computing cluster.
The code for these experiments is available at the GitHub repository \texttt{aaamini/nsbm}~\citep{Josephs_NSBM_for_simultaneously_2023}.

An overview of our findings is as follows:
\begin{enumerate}[(1), itemsep=1ex]
    \item There is a trade-off in performance with respect to clustering networks and nodes.
    \item (IBG) performs the best clustering networks, followed by (CG).
    \item (CG) performs the best clustering nodes, followed by (G).
    \item (IBG) typically outperforms (BG), especially for larger networks.
    \item (CG) and (G) are more stable than (BG) and (IBG)
    \item Overall, (CG) performs the best.
\end{enumerate}

\subsection{Simulation setting}

We begin by defining a general data-generating mechanism that we will use throughout our simulations to sample a collection of networks.

For each network $j = 1, \ldots, J$, we begin by sampling its class membership $z_j \sim \{1, \ldots, K\}$.
For each network class, we then sample $\xib_{k}^* \sim \unif(\{1, \ldots, L(k)\})^{\otimes n}$.
To introduce community heterogeneity, 
we first let $\xib_j = \xib_{z_j}^*$ and then for each coordinate of $\xib_j$,  independently and with probability $\tau$,  resample the coordinate from $\unif(\{1, \ldots, L(k)\})$. This creates a Markov perturbation of community labels within a network cluster, where each node retains its class level community with probability $1-\tau$ or gets a new label with probability $\tau.$
Consequently, although the nodes are aligned, the community membership of a given node may change within a network class when $\tau > 0$.
When $\tau = 0$, the community structure is identical for networks in the same class (the labeled case), but the community heterogeneity increases as $\tau$ increases.
Finally, we sample the networks
\begin{equation}\label{eq:sim_A}
    A^j \mid \etab,\, \xib_{j}, z_j \;\sim\; \sbm \Bigl(\xib_j, \alpha_j  \etab_{z_j} \Bigr) \enskip ,
\end{equation}
where $\etab_{z_j}$ is the connectivity matrix for class $z_j$ and $\alpha_j$ is a scaling parameter.

\subsection{Convergence analysis}\label{sec:run_time}

We being by comparing the Gibbs samplers from Section~\ref{sec:Gibbs}.
For $k = 1, \ldots, K$ and $\gamma \in [0, 1]$, we let
\begin{equation}\label{eq:sim_eta}
    \etab_k = (1-\gamma) \cdot I_{L(k)} + \gamma \cdot U_{L(k)} \enskip ,
\end{equation}
where $I_n$ is the $n \times n$ identity matrix of and $U_n$ is a random symmetric  $n \times n$ matrix, with entries uniformly distributed in $[0,1]$.
For $\gamma = 0$, we have a perfectly assortative stochastic block model, which becomes less assortative as $\gamma$ increases to~1.
Therefore, $\gamma$ controls the difficulty of clustering the networks.
We let
\begin{equation}\label{eq:ead}
    \alpha_j = \frac{\lambda}{\text{ead}(n_j, \etab_{z_j})}
    \enskip ,
\end{equation}
where $\text{ead}(n_j, \etab_{z_j})$ is the expected average degree (EAD) of an SBM with $n_j$ nodes and connectivity matrix $\etab_{z_j}$.
The scaling of $\etab_{z_j}$ by $\alpha_j$ is done to obtain an SBM with an EAD of $\lambda$, a parameter that can then be used to control the sparsity of the network.
As $\lambda$ increases, the networks become more sparse making community detection more difficult.

\begin{figure}[t]
    \centering
    \includegraphics[width=\textwidth]{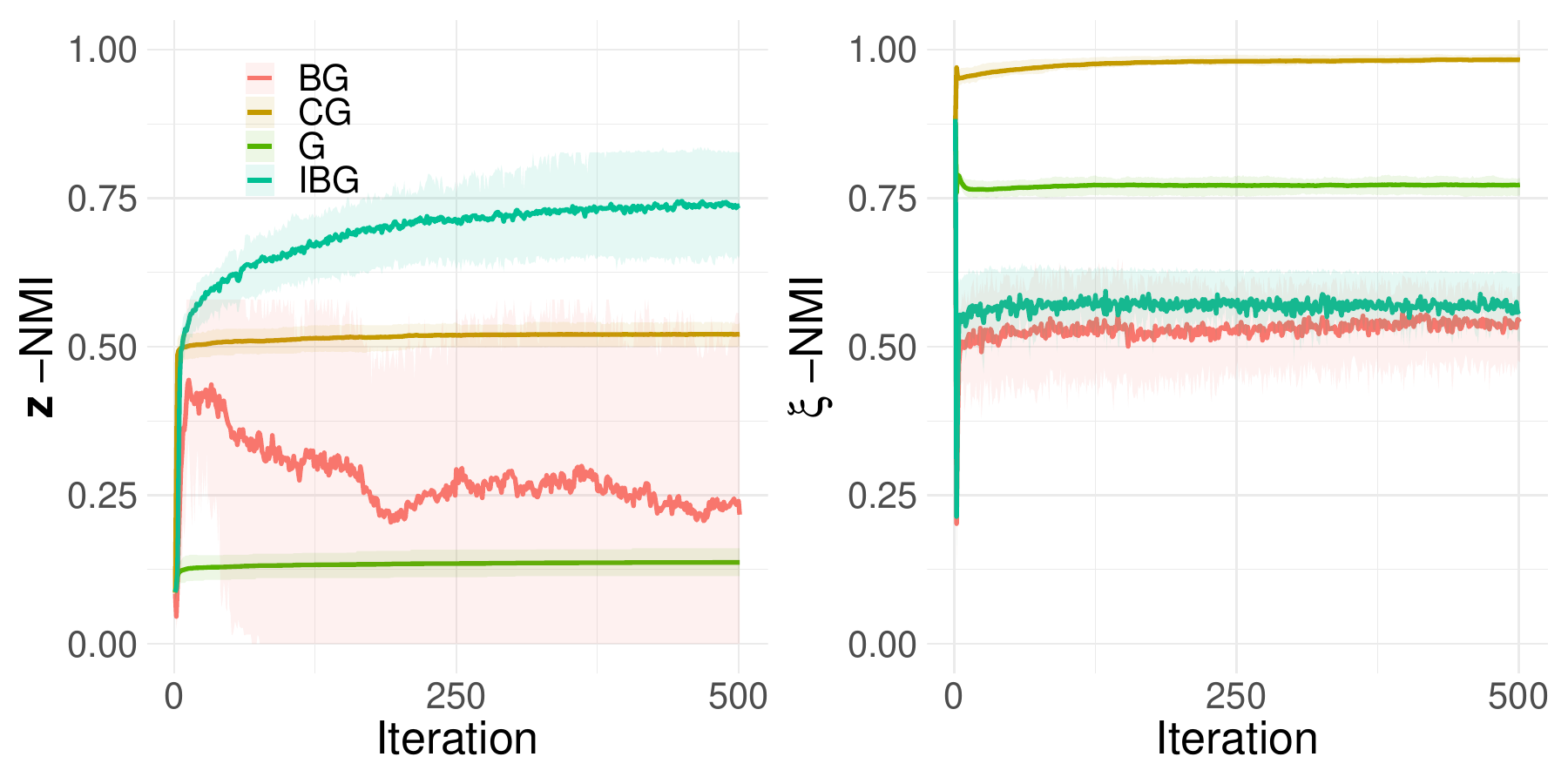}
    \caption{Cluster performance of our four Gibbs algorithms. $\zb$-NMI (left) measures how well we are clustering the network objects and $\xib$-NMI (right) measures how well we are performing community detection on the nodes. The bands are 50\% quantile regions based on 100 experiments.}
    \label{fig:run_time}
\end{figure}

We take $J = 60$ networks on $n = 200$ nodes each with $\gamma = 0.1$ and $\lambda = 30$.
We also let $\tau = 0$ so that we are working in the labeled case, although as we will see later, NSBM is not affected by $\tau$.
We sample evenly from $K = 3$ classes with $L =$ 2, 3, and 5 communities.
We repeat each experiment 100 times and plot the mean $\zb$-NMI and $\xib$-NMI along with the interquartile range in Figure~\ref{fig:run_time}.

The first thing we see is that each of the Gibbs samplers ``converges" quickly.
This highlights how our samplers are able to efficiently find good clusters after just a few iterations.
Focusing on the early part of the chain, we observe an interesting phenomenon that highlights the trade-off between learning $\bm z$ and learning $\bm \xi$.
In particular, Figure~\ref{fig:run_time} illustrates that NSBM provides improved community detection when combining the networks rather than performing community detection individually.
We see that $\bm\xi$-NMI starts high (due to our DP-SBM initializations), goes down while the model is learning $\bm z$, and then comes back up, in some cases higher than the initial values.

Next, we see that (IBG) outperforms (BG) on $\zb$-NMI and vice-versa for $\xib$-NMI despite only differing in the order of their exact joint sampling of $\zb$ and $\xib$.
\citet{van2008partially} suggest that incompatibility can actually improve mixing in partially collapsed Gibbs samplers at the risk of sacrificing the correct correlation structure between parameters.
In terms of clustering performance, our results indicate that this sacrifice may be worthwhile depending on the parameter of interest.

Finally, we see that although (IBG) and (CG) perform the best, on average, on $\zb$ and $\xib$, respectively, there is considerable variability in the results.
This is especially prominent with (BG)and (IBG), as the interquartile band demonstrates, compared to the more stable (CG) and (G).
One interesting finding that relates to this variability is that we observed $\zb$ collapse to a single class, which results in $\zb$-NMI of 0.
This may be due to the previously unmentioned non-identifiability in NDP mixture models that we introduced in Section~\ref{sec:ndp_and_nsbm}.
Specifically, since the number of classes and clusters can take any value, a single class can fully capture the data with a growing number of clusters.
In this example, that means we cannot distinguish between $K = 3$ classes with $L = 2, 3$, and 5 communities compared to a single class with $L = 10$ communities.

\subsection{Scaling analysis}

In this simulation, we keep the same setup from Section~\ref{sec:run_time}, but we vary $n$ to see how the samplers scale with the network order.
However, for the scaling parameter in Equation~\eqref{eq:ead}, we let $\lambda = n / 10$ so that the average degree grows with the network.
Throughout, we take $J = 30$ and $\gamma = 0.4$, and we vary $n$ from $20$ to $500$.

\begin{figure}[t]
    \centering
    \includegraphics[width=\textwidth]{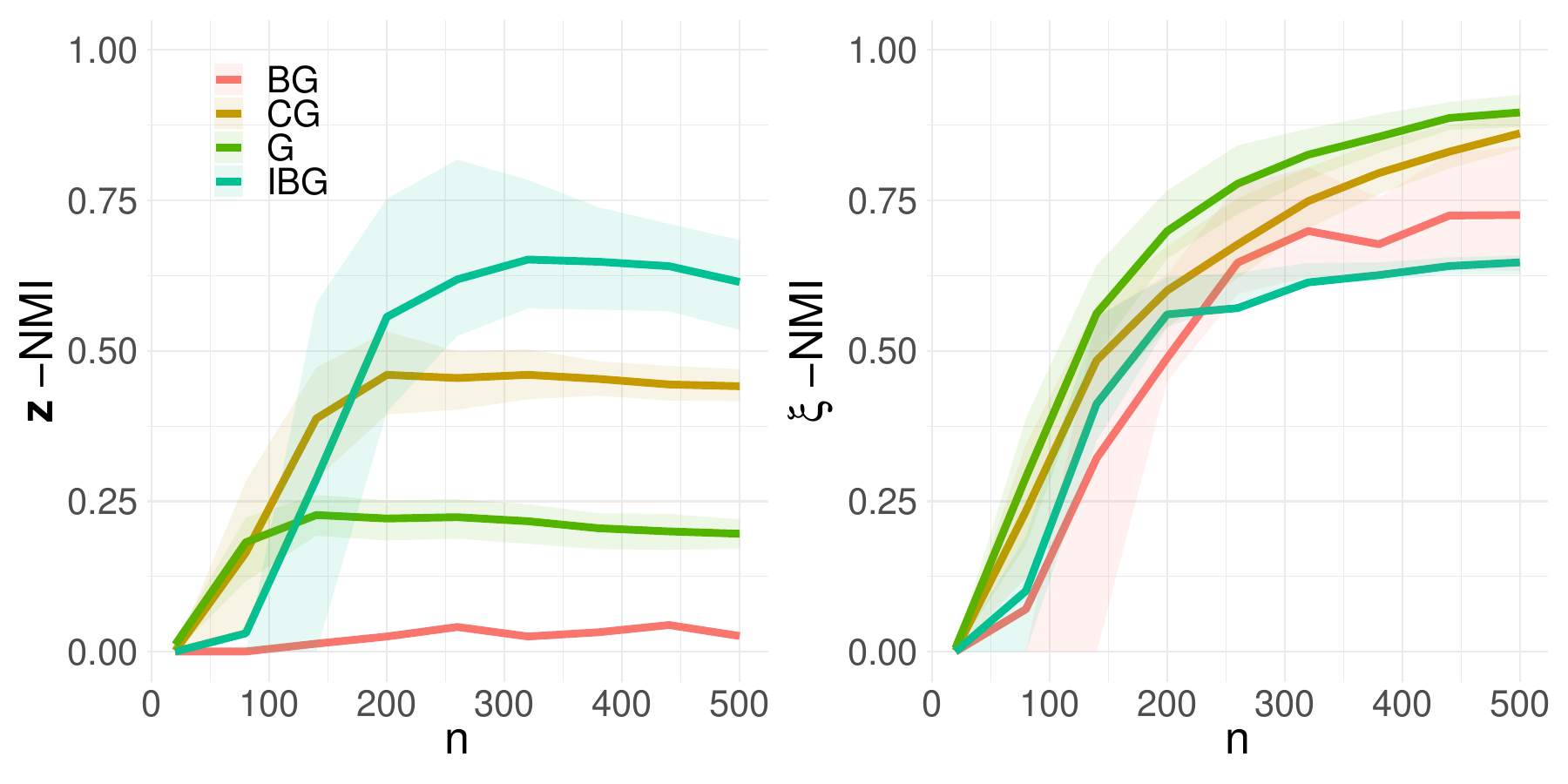}
    \caption{Varying $n$ with $\lambda = n/10$ and $\gamma = 0.4$.}
    \label{fig:scaling}
\end{figure}

Figure~\ref{fig:scaling} shows the results.
We see that $\xib$-NMI increases as the networks grow for all of the methods, which is what we expect for community detection.
As before, we see that (G) and (CG) outperform (BG) and (IGB).
We also see a sharp increase in $\zb$-NMI for (G), (CG), and (IBG), with (IBG) performing the best for large networks.
Interestingly, $\zb$-NMI is stable for (BG), which can be explained by noting that blocking requires marginalization over $nL + n$ terms, hence its chain might be slower to converge.
In contrast, the improvement of (IBG) provides further evidence that incompatibility can improve mixing.
Similarly, by marginalizing out $\etab$, the number of parameters that (CG) samples grows linearly with $n$ unlike with (G).
Another interesting finding is that $\zb$-NMI eventually stabilizes for all of the samplers.
There may be an underlying trade-off between the number of nodes $n$ and the number of networks $J$ but this relationship needs to be examined further.

\subsection{Community heterogeneity analysis}\label{sec:sim_heterogeneity}

In this simulation, we vary $\tau \in [0, 1]$.
Throughout, we take $J = 20$ and $n = 200$.
We also fix $\gamma = 0.2$ and $\lambda = 25$ from Equations~\eqref{eq:sim_eta} and \eqref{eq:ead}, respectively.
We omit BG to declutter the results due to its poor performance in the previous simulations.

\begin{figure}[t]
    \centering
    \includegraphics[width=\textwidth]{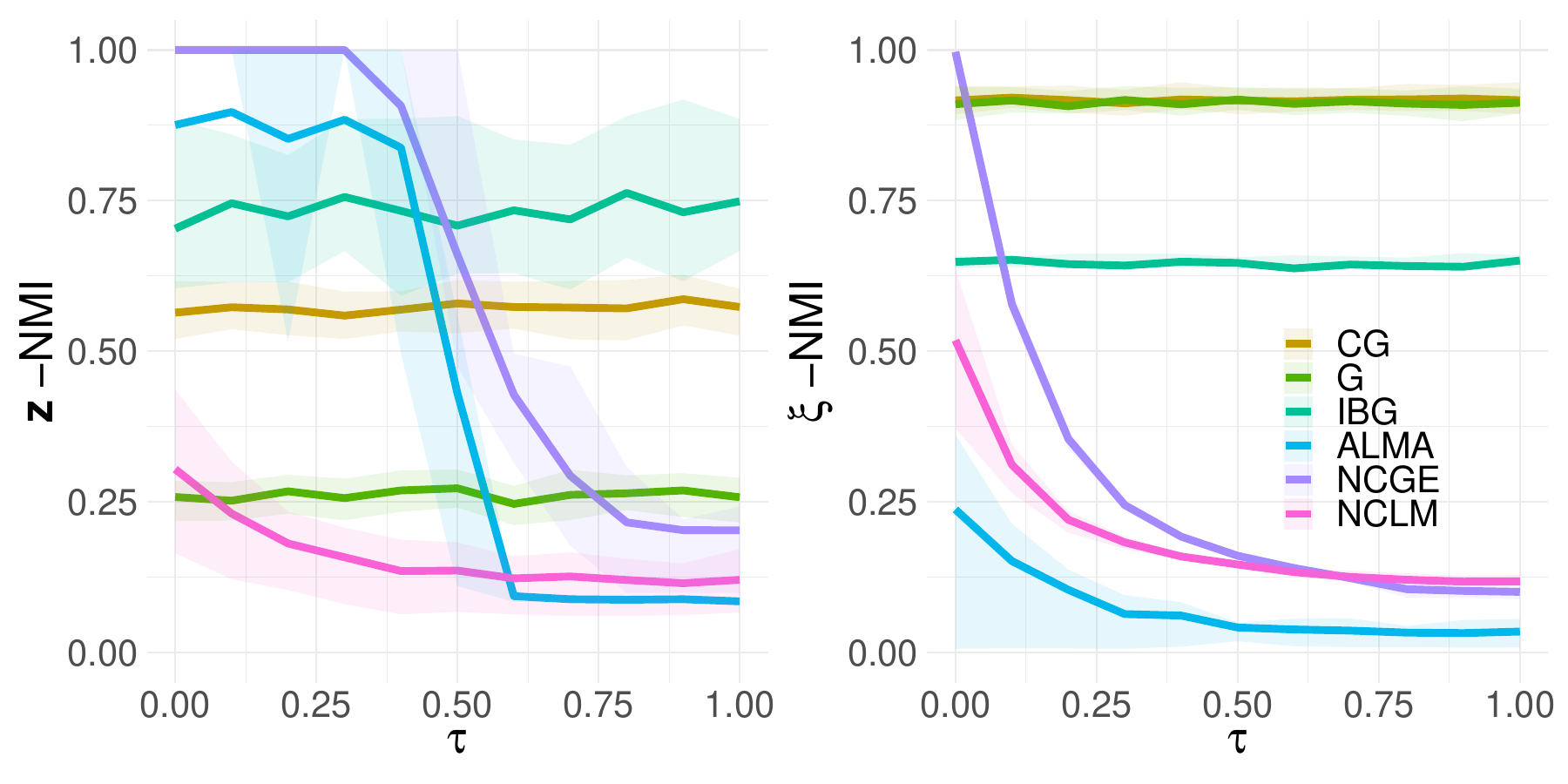}
    \caption{Varying $\tau$ for fixed $\gamma = 0.2$ and $\lambda = 25$.}
    \label{fig:transprob}
\end{figure}

Figure~\ref{fig:transprob} shows the results.
We see that the NSBM samplers are robust to community heterogeneity as $\zb$-NMI and $\xib$-NMI are stable for all values of $\tau$.
On the other hand, NCGE and ALMA exhibit a sharp phase transition for $\tau \geq 0.5$: with low community heterogeneity, NCGE and ALMA perform the best, whereas with high community heterogeneity, they perform uniformly worse than NSBM even though we provide them the true number of communities.
It is not surprising that NCGE and ALMA outperform NSBM for low $\tau$, because they use the additional information encoded in the alignment of the nodes, whereas NSBM does not utilize the alignment at all.
Lastly, NCLM has low $\zb$-NMI for all values of $\tau$.
Importantly, we see that only the NSBM samplers exhibit good community detection performance, while NCGE, NCLM, and ALMA show decreasing $\xib$-NMI as $\tau$ increases.

\subsection{Non-assortative networks}

\begin{figure}[t]
    \centering
    \includegraphics[width=\textwidth]{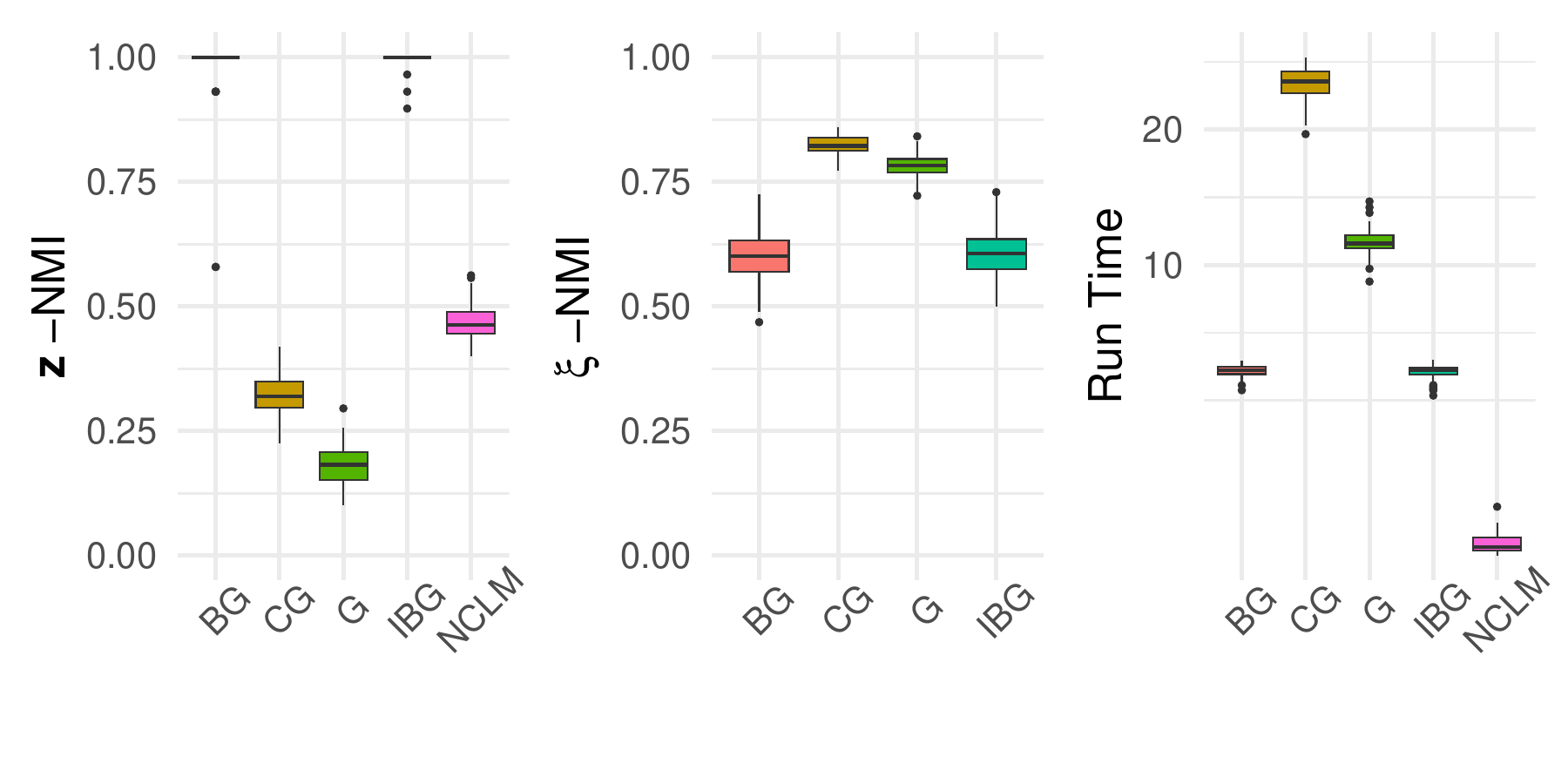}
    \caption{Simulated multilayer personality-friendship networks.}
    \label{fig:HSBM}
\end{figure}

Here, we test our model on non-assortative networks based on multilayer personality-friendship networks from \citet{amini2022hierarchical}.
In this setting, we suppose there are three types of schools that have different interaction frequencies between students.
In each school, students belong to one of three personalty types: extrovert, ambivert, or introvert.
The probabilities of interaction between students of various personality types are given as
\begin{equation*}
\renewcommand{\arraystretch}{0.6} 
    \etab_1 = \begin{pmatrix}
        .9 & .75 & .5 \\
        .75 & .6 & .25 \\
        .5 & .25 & .1
    \end{pmatrix}, \quad \etab_2 = \begin{pmatrix}
        .8 & .1 & .3 \\
        .1 & .9 & .2 \\
        .3 & .2 & .7
    \end{pmatrix}, \quad \etab_3 = \begin{pmatrix}
        .1 & .4 & .6 \\
        .4 & .3 & .1 \\
        .6 & .1 & .5
    \end{pmatrix} ,
\renewcommand{\arraystretch}{1}     
\end{equation*}
where $\etab_k$ represents the probabilities for school type $k$.
We see that for school type~1, extroverts interact most often with other extroverts, but are still marginally more likely to interact with any student, whereas for school type~2, the students mix assortatively within their personality type.
The third school type is a mix of the first two schools: ambiverts and introverts mix assortatively, whereas extroverts prefer to mix with non-extroverts.
The proportion of students belonging to these personality types is given in Table~\ref{tab:hsbm}.
We sample 40 social networks for each school ($J = 120$) and each school has $n \sim \unif(20, 100)$ students.
In this setting, we take the networks to be unlabeled ($\tau = 1$), which reflects realistic anonymity and differing node sets in real social networks.
Therefore, we only include NCLM as a comparison and only in its performance  clustering the networks.
The results for 100 replicates are given in Figure~\ref{fig:HSBM}.

\begin{table}[t]
\centering
\begin{tabular}{@{}cccc@{}}
\toprule
\textbf{School} & \textbf{Extrovert} & \textbf{Ambivert} & \textbf{Introvert} \\
\midrule
\textbf{1} & 40 & 35 & 25 \\
\textbf{2} & 70 & 15 & 15 \\
\textbf{3} & 20 & 40 & 40 \\
\bottomrule
\end{tabular}
\caption{Percentage of students belonging to the three personality types for each of the school types.}
\label{tab:hsbm}
\end{table}

As in the previous simulations, (CG) and (G) perform the best on $\xib$.
(BG) and (IBG) both perform very well on $\zb$, dominating NCLM.
In this case, (BG) performs similar to its incompatible version, providing evidence that (BG) performs well on small networks.

\section{Real data analysis}\label{sec:data}

To showcase the flexibility of NSBM, we consider two real datasets.
First, we apply it to an aggregated dataset consisting of five different types of social network from \citet{oettershagen2020temporal}: high school, Facebook, Tumblr, MIT, DBLP.
In total, there are over 2300 networks from these 5 datasets ranging from $n =$ 20 to 100.
This dataset accurately reflects the difficulty in comparing social networks since they are anonymous and also have a different number of nodes per network even within network type.

In order to resemble the more common setting in which only a few social networks are observed, we perform an \textit{in silico} study by sampling $J \sim \unif(20, 100)$ per experiment.
We compare the results of NSBM to NCLM, the only other method that works for unlabeled networks, on 100 different experiments.
For the comparison, the true classes are the social network type (high school, Facebook, Tumblr, MIT, DBLP).
However, as the networks are anonymous, we do not have ground-truth node labels.
Therefore, we only compare NSBM and NCLM on $\zb$-NMI and not on $\xib$-NMI.
The results in Table~\ref{tab:real} match what we found in the simulation on the multilayer personality-friendship networks.
That is, (IBG) and (BG) perform the best.
(IBG) significantly outperforms NCLM even though NCLM is given the true number of classes.

We also apply NSBM to character-interaction networks from popular films and television series \citep{david2020networkdata}.
We consider $J = 15$ networks: 6 from the original and sequel Star Wars trilogies, 7 from the Game of Thrones television series, and 2 Lord of the Rings films.
For each network, the nodes represent characters and an edge exists between two characters if they share a scene together.
The number of nodes ranges from $n = 20$ to 172.
In this case, we can assess the clustering performance of NSBM on both network and node labels.
For the network labels, we let each network belong to its franchise (Star Wars, Game of Thrones, and Lord of the Rings).
Since we have character names, we can also assign node labels.
For the characters in Star Wars, we assign them to an ``affiliation" based on their Wookiepedia entry (\url{https://starwars.fandom.com/}).
The labels are: Rebel Alliance, Galactic Republic, Galactic Empire, Confederacy of Independent Systems, Jedi Order, and Sith Order.
Of the 91 unique characters included in the six films, 25 do not belong to one of these affiliations.
Therefore, we discard them from our cluster performance evaluation, but note that we leave them in as nodes in the networks.
We do not assign labels to the Game of Thrones networks because there are 400 unique characters, which we leave for future researchers.

The results are shown in Table~\ref{tab:real}.
Again, we see that (IBG) performs the best on $\zb$-NMI.
We also see that (BG), (CG), and (G) perform the best on $\xib$-NMI.
We note that the low $\xib$-NMI performance can, at least partially, be explained by the arbitrary affiliations we assigned the characters.
In particular, we used the last reported affiliation for convenience, but an affiliation that varies with time or film may be more appropriate.
Alternatively, since many of the characters belong to several affiliations, a more appropriate model might be a mixed-membership stochastic block model.
Finally, some of the affiliations are more similar than others, for example Galactic Empire and Galactic Republic, but NMI does not weigh these mismatches differently than any random pair.
A better understanding of the affiliation hierarchy could yield a better performance metric.

\begin{table}[t]
\centering
\begin{adjustbox}{max width=\textwidth}
\begin{tabular}{lP{1.8cm}P{1.8cm}P{1.8cm}P{1.8cm}P{1.8cm}}
\toprule
\textbf{Dataset} & \textbf{BG} & \textbf{CG} & \textbf{G} & \textbf{IBG} & \textbf{NCLM} \\ \midrule
\cite{oettershagen2020temporal} $(\zb)$ & 0.50 \newline(0.45-0.58) & 0.33	(0.28-0.36) & 0.21 (0.16-0.33) & \textbf{0.63} (0.56-0.70) & 0.47 (0.44-0.51) \\ \midrule
\cite{david2020networkdata} $(\zb)$ & 0 & 0.31 & 0.30 & \textbf{0.70} & 0.36 \\ \midrule
\cite{david2020networkdata} $(\xib)$ & \textbf{0.37} & 0.22 & 0.23 & 0.08 & NA \\ \bottomrule
\end{tabular}
\end{adjustbox}

\medskip
\caption{NMI (interquartile range) comparison of NSBM and NCLM.}
\label{tab:real}
\end{table}

\section{Conclusion and discussions}\label{sec:conclusion}

In this work, we have introduced the nested stochastic block model (NSBM).
By novelly employing the nested Dirichlet process (NDP) prior \citep{rodriguez2008nested}, NSBM has the flexibility to cluster unlabeled networks while simultaneously performing community detection.
Importantly, NSBM also learns the number of classes at the network and node levels.
NSBM is not a straightforward application of an NDP mixture model because network edges represent pairwise interactions between nodes and thus violates the original i.i.d. setting.
This further manifests as the inability to use the Gibbs sampler originally proposed in \citet{rodriguez2008nested}.
As a consequence, we proposed four different Gibbs samplers that are shown to have various strengths in our numerical results.

Importantly, although three of the samplers have the same (correct) posterior, we observe that they seem to converge to different clustering solutions.
In particular, (IBG) tends to posterior modes that correspond to accurate network clustering, while (G) converges to accurate node clustering.
Since (CG) performs well in both tasks, we conclude that it is the best sampler, but it remains open why these three samplers, with identical stationary distributions, perform differently.
One hypothesis is that the non-identifibaility of the NDP likelihood coupled with a weak label prior leads to a truly multimodal posterior with regions of very low probability in between, trapping each sampler in the vicinity of a different mode.

We also observe that although (IBG) is incompatible, it outperforms (BG) on the clustering tasks.
\citet{van2008partially} suggest that incompatibility, while breaking the parameter dependency and creating a stationary distribution that differs from the true posterior, can improve mixing.
Since mixing is a known challenge for DP mixture models \citep{neal2000markov, hastie2015sampling}, this may be worthwhile.
However, more theory is needed to understand this trade-off.
In particular, the relationship between the marginal posterior distributions and the global posterior distribution, and how they relate to clustering performance, needs to be studied.

One can look at other samplers besides the four we considered.
As mentioned in the discussion of (BG), computing $p(z_j = k \given E)$ is difficult. It boils down to computing sums of the form \begin{align}\label{eq:hard:sum}
\sum_{x_1,x_2,\dots,x_{n_j }} \prod_{s <t} f_{st}(x_s, x_t) \enskip , 
\end{align}
where $f_{st}(x_s,x_t) = \eta_{x_s x_t k}^{A_{st}^j} (1-\eta_{x_s x_t k})^{A_{st}^j}$.
Loopy belief propagation (BP), a message-passing algorithm, can be used to approximately compute the sum.
Still, this requires passing messages over the complete graph. Ideas from~\citet{decelle2011asymptotic} can be used to further approximate by passing messages only over edges of $A^j$, leading to a speed up for sparse networks. We can also use a Gibbs sampler to compute~\eqref{eq:hard:sum}, leading to Gibbs-within-Gibbs samplers. These samplers are interesting to implement and study; however, for practical use, they are mostly infeasible due to the computational cost of calculating $p(z_j = k \given E)$ in each step.

One important discovery that we made in our simulations is the phenomena that chains occasionally collapse into a single class for $\zb$.
We experienced this for all of the samplers, though most notably with (BG), and it only occurs some of the times for some of the settings.
We believe this is due to an inherent non-identifiability in NDP mixture models, which has not been previously discussed.
This issue is of independent interest and should be investigated further.
It is also unclear why this issue did not occur in the original NDP simulations.


Finally, there are several natural modifications that could be made to NSBM including extensions to undirected networks, as well as to degree-corrected and mixed-membership stochastic block models.
We leave these extensions to future researchers.


\section*{Acknowledgments}
NJ was partially supported by NIH/NICHD grant 1DP2HD091799-01.
AA was supported by NSF grant DMS-1945667. 
LL was supported by NSF grants DMS 2113642 and DMS 1654579.

\printbibliography

@software{Josephs_NSBM_for_simultaneously_2023,
author = {Josephs, Nathaniel and Amini, Arash A. and Paez, Marina and Lin, Lizhen},
month = jul,
title = {{NSBM for simultaneously clustering networks and nodes}},
url = {https://github.com/aaamini/nsbm},
year = {2023}
}

@article{chen2022global,
  title={Global and individualized community detection in inhomogeneous multilayer networks},
  author={Chen, Shuxiao and Liu, Sifan and Ma, Zongming},
  journal={The Annals of Statistics},
  volume={50},
  number={5},
  pages={2664--2693},
  year={2022},
  publisher={Institute of Mathematical Statistics}
}

@article{abbe2017community,
  title={Community detection and stochastic block models: recent developments},
  author={Abbe, Emmanuel},
  journal={The Journal of Machine Learning Research},
  volume={18},
  number={1},
  pages={6446--6531},
  year={2017},
  publisher={JMLR. org}
}

@article{decelle2011asymptotic,
  title={Asymptotic analysis of the stochastic block model for modular networks and its algorithmic applications},
  author={Decelle, Aurelien and Krzakala, Florent and Moore, Cristopher and Zdeborov{\'a}, Lenka},
  journal={Physical review E},
  volume={84},
  number={6},
  pages={066106},
  year={2011},
  publisher={APS}
}

@article{arroyo2021inference,
  title={Inference for multiple heterogeneous networks with a common invariant subspace},
  author={Arroyo, Jes{\'u}s and Athreya, Avanti and Cape, Joshua and Chen, Guodong and Priebe, Carey E and Vogelstein, Joshua T},
  journal={The Journal of Machine Learning Research},
  volume={22},
  number={1},
  pages={6303--6351},
  year={2021},
  publisher={JMLRORG}
}

@article{holland1981exponential,
  title={An exponential family of probability distributions for directed graphs},
  author={Holland, Paul W and Leinhardt, Samuel},
  journal={Journal of the american Statistical association},
  volume={76},
  number={373},
  pages={33--50},
  year={1981},
  publisher={Taylor \& Francis}
}

@article{le2018estimating,
  title={Estimating a network from multiple noisy realizations},
  author={Le, Can M and Levin, Keith and Levina, Elizaveta},
  year={2018}
}

@article{lei2020consistent,
  title={Consistent community detection in multi-layer network data},
  author={Lei, Jing and Chen, Kehui and Lynch, Brian},
  journal={Biometrika},
  volume={107},
  number={1},
  pages={61--73},
  year={2020},
  publisher={Oxford University Press}
}

@article{usvt,
author = {Sourav Chatterjee},
title = {{Matrix estimation by Universal Singular Value Thresholding}},
volume = {43},
journal = {The Annals of Statistics},
number = {1},
publisher = {Institute of Mathematical Statistics},
pages = {177 -- 214},
year = {2015},
doi = {10.1214/14-AOS1272},
URL = {https://doi.org/10.1214/14-AOS1272}
}

@ARTICLE{zhang_et_al,
   author = {{Zhang}, Y. and {Levina}, E. and {Zhu}, J.},
    title = "{Estimating network edge probabilities by neighborhood smoothing}",
  journal = {ArXiv e-prints},
archivePrefix = "arXiv",
   eprint = {1509.08588},
 primaryClass = "stat.ML",
     year = 2015,
    month = sep,
   adsurl = {http://adsabs.harvard.edu/abs/2015arXiv150908588Z}
}

@article{mukherjee2017clustering,
  title={On clustering network-valued data},
  author={Mukherjee, S and Sarkar, P and Lin, L},
  journal={Advances in neural information processing systems},
  year={2017}
}

@article{ginestet2017hypothesis,
  title={Hypothesis testing for network data in functional neuroimaging},
  author={Ginestet, Cedric E and Li, Jun and Balachandran, Prakash and Rosenberg, Steven and Kolaczyk, Eric D},
  journal={The Annals of Applied Statistics},
  pages={725--750},
  year={2017},
  publisher={JSTOR}
}

@article{kolaczyk2020averages,
  title={Averages of unlabeled networks: Geometric characterization and asymptotic behavior},
  author={Kolaczyk, Eric D and Lin, Lizhen and Rosenberg, Steven and Walters, Jackson and Xu, Jie},
  journal={The Annals of Statistics},
  volume={48},
  number={1},
  pages={514--538},
  year={2020},
  publisher={Institute of Mathematical Statistics}
}

@article{chen2019hypothesis,
  title={Hypothesis testing for populations of networks},
  author={ Chen, L and  Zhou, J and   Lin, L},
  journal={arXiv preprint arXiv:1911.03783},
  year={2019}
}

@article{chen2020spectral,
  title={A spectral-based framework for hypothesis testing in populations of networks},
  author={Chen, Li and Josephs, Nathaniel and Lin, Lizhen and Zhou, Jie and Kolaczyk, Eric D},
  journal={Statistics Sinica},
  year={2023}
}

@article{relion2019network,
  title={Network classification with applications to brain connectomics},
  author={Arroyo Reli{\'o}n, Jes{\'u}s D and Kessler, Daniel and Levina, Elizaveta and Taylor, Stephan F},
  journal={The annals of applied statistics},
  volume={13},
  number={3},
  pages={1648},
  year={2019},
  publisher={NIH Public Access}
}

@article{orbanz2014bayesian,
  title={Bayesian models of graphs, arrays and other exchangeable random structures},
  author={Orbanz, Peter and Roy, Daniel M},
  journal={IEEE transactions on pattern analysis and machine intelligence},
  volume={37},
  number={2},
  pages={437--461},
  year={2014},
  publisher={IEEE}
}

@article{josephs2023bayesian,
  title={Bayesian classification, anomaly detection, and survival analysis using network inputs with application to the microbiome},
  author={Josephs, Nathaniel and Lin, Lizhen and Rosenberg, Steven and Kolaczyk, Eric D},
  journal={The Annals of Applied Statistics},
  volume={17},
  number={1},
  pages={199--224},
  year={2023},
  publisher={Institute of Mathematical Statistics}
}

@inproceedings{josephs2021network,
  title={Network recovery from unlabeled noisy samples},
  author={Josephs, Nathaniel and Li, Wenrui and Kolaczyk, Eric D},
  booktitle={2021 55th Asilomar Conference on Signals, Systems, and Computers},
  pages={1268--1273},
  organization={IEEE},
  year={2021}
}

@article{shen2022bayesian,
  title={Bayesian community detection for networks with covariates},
  author={Shen, Luyi and Amini, Arash and Josephs, Nathaniel and Lin, Lizhen},
  journal={arXiv preprint arXiv:2203.02090},
  year={2022}
}

@article{rodriguez2008nested,
  title={The nested Dirichlet process},
  author={Rodriguez, Abel and Dunson, David B and Gelfand, Alan E},
  journal={Journal of the American Statistical Association},
  volume={103},
  number={483},
  pages={1131--1154},
  year={2008},
  publisher={Taylor \& Francis}
}

@article{sethuraman1994constructive,
  title={A constructive definition of Dirichlet priors},
  author={Sethuraman, Jayaram},
  journal={Statistica sinica},
  pages={639--650},
  year={1994},
  publisher={JSTOR}
}

@techreport{pitman2002combinatorial,
  title={Combinatorial stochastic processes},
  author={Pitman, Jim and others},
  year={2002},
  institution={Technical Report 621, Dept. Statistics, UC Berkeley, 2002. Lecture notes for~…}
}

@article{paul2020random,
author = {Subhadeep Paul and Yuguo Chen},
title = {{A random effects stochastic block model for joint community detection in multiple networks with applications to neuroimaging}},
volume = {14},
journal = {The Annals of Applied Statistics},
number = {2},
publisher = {Institute of Mathematical Statistics},
pages = {993 -- 1029},
year = {2020},
doi = {10.1214/20-AOAS1339},
URL = {https://doi.org/10.1214/20-AOAS1339}
}

@article{arroyo2020simultaneous,
  title={Simultaneous prediction and community detection for networks with application to neuroimaging},
  author={Arroyo, Jes{\'u}s and Levina, Elizaveta},
  journal={arXiv preprint arXiv:2002.01645},
  year={2020}
}

@article{fortunato2010community,
  title={Community detection in graphs},
  author={Fortunato, Santo},
  journal={Physics reports},
  volume={486},
  number={3-5},
  pages={75--174},
  year={2010},
  publisher={Elsevier}
}

@article{young2022clustering,
  title={Clustering of heterogeneous populations of networks},
  author={Young, Jean-Gabriel and Kirkley, Alec and Newman, MEJ},
  journal={Physical Review E},
  volume={105},
  number={1},
  pages={014312},
  year={2022},
  publisher={APS}
}

@article{mantziou2021bayesian,
  title={Bayesian model-based clustering for multiple network data},
  author={Mantziou, Anastasia and Lunagomez, Simon and Mitra, Robin},
  journal={arXiv preprint arXiv:2107.03431},
  year={2021}
}

@article{fan2022alma,
  title={ALMA: Alternating Minimization Algorithm for Clustering Mixture Multilayer Network},
  author={Fan, Xing and Pensky, Marianna and Yu, Feng and Zhang, Teng},
  journal={Journal of Machine Learning Research},
  volume={23},
  number={330},
  pages={1--46},
  year={2022}
}

@article{signorelli2020model,
  title={Model-based clustering for populations of networks},
  author={Signorelli, Mirko and Wit, Ernst C},
  journal={Statistical Modelling},
  volume={20},
  number={1},
  pages={9--29},
  year={2020},
  publisher={SAGE Publications Sage India: New Delhi, India}
}

@article{van2008partially,
  title={Partially collapsed Gibbs samplers: Theory and methods},
  author={Van Dyk, David A and Park, Taeyoung},
  journal={Journal of the American Statistical Association},
  volume={103},
  number={482},
  pages={790--796},
  year={2008},
  publisher={Taylor \& Francis}
}

@inproceedings{pedarsani2011privacy,
  title={On the privacy of anonymized networks},
  author={Pedarsani, Pedram and Grossglauser, Matthias},
  booktitle={Proceedings of the 17th ACM SIGKDD international conference on Knowledge discovery and data mining},
  pages={1235--1243},
  year={2011}
}

@article{newman2016structure,
  title={Structure and inference in annotated networks},
  author={Newman, Mark EJ and Clauset, Aaron},
  journal={Nature communications},
  volume={7},
  number={1},
  pages={1--11},
  year={2016},
  publisher={Nature Publishing Group}
}

@inproceedings{kemp, author = {Kemp, Charles and Tenenbaum, Joshua B. and Griffiths, Thomas L. and Yamada, Takeshi and Ueda, Naonori}, title = {Learning Systems of Concepts with an Infinite Relational Model}, year = {2006}, isbn = {9781577352815}, publisher = {AAAI Press}, booktitle = {Proceedings of the 21st National Conference on Artificial Intelligence - Volume 1}, pages = {381–388}, numpages = {8}, location = {Boston, Massachusetts}, series = {AAAI'06} }

@inproceedings{zhou2015infinite,
  title={Infinite edge partition models for overlapping community detection and link prediction},
  author={Zhou, Mingyuan},
  booktitle={Artificial intelligence and statistics},
  pages={1135--1143},
  year={2015},
  organization={PMLR}
}

@InProceedings{pmlr-v70-zhao17a,
  title = 	 {Leveraging Node Attributes for Incomplete Relational Data},
  author =       {He Zhao and Lan Du and Wray Buntine},
  booktitle = 	 {Proceedings of the 34th International Conference on Machine Learning},
  pages = 	 {4072--4081},
  year = 	 {2017},
  editor = 	 {Precup, Doina and Teh, Yee Whye},
  volume = 	 {70},
  series = 	 {Proceedings of Machine Learning Research},
  month = 	 {06--11 Aug},
  publisher =    {PMLR},
  url = 	 {https://proceedings.mlr.press/v70/zhao17a.html}
}

@inproceedings{meta,
  author    = {Dae Il Kim and
               Michael C. Hughes and
               Erik B. Sudderth},
  title     = {The Nonparametric Metadata Dependent Relational Model},
  booktitle = {Proceedings of the 29th International Conference on Machine Learning,
               {ICML} 2012, Edinburgh, Scotland, UK, June 26 - July 1, 2012},
  publisher = {icml.cc / Omnipress},
  year      = {2012},
  url       = {http://icml.cc/2012/papers/771.pdf},
  bibsource = {dblp computer science bibliography, https://dblp.org}
}

@article{amini2022hierarchical,
  title={Hierarchical stochastic block model for community detection in multiplex networks},
  author={Amini, Arash and Paez, Marina and Lin, Lizhen},
  journal={Bayesian Analysis},
  volume={1},
  number={1},
  pages={1--27},
  year={2022},
  publisher={International Society for Bayesian Analysis}
}

@article{hastie2015sampling,
  title={Sampling from Dirichlet process mixture models with unknown concentration parameter: mixing issues in large data implementations},
  author={Hastie, David I and Liverani, Silvia and Richardson, Sylvia},
  journal={Statistics and computing},
  volume={25},
  number={5},
  pages={1023--1037},
  year={2015},
  publisher={Springer}
}

@article{neal2000markov,
  title={Markov chain sampling methods for Dirichlet process mixture models},
  author={Neal, Radford M},
  journal={Journal of computational and graphical statistics},
  volume={9},
  number={2},
  pages={249--265},
  year={2000},
  publisher={Taylor \& Francis}
}

@article{lunagomez2021modeling,
  title={Modeling network populations via graph distances},
  author={Lunag{\'o}mez, Sim{\'o}n and Olhede, Sofia C and Wolfe, Patrick J},
  journal={Journal of the American Statistical Association},
  volume={116},
  number={536},
  pages={2023--2040},
  year={2021},
  publisher={Taylor \& Francis}
}

@article{jing2021community,
  title={Community detection on mixture multilayer networks via regularized tensor decomposition},
  author={Jing, Bing-Yi and Li, Ting and Lyu, Zhongyuan and Xia, Dong},
  journal={The Annals of Statistics},
  volume={49},
  number={6},
  pages={3181--3205},
  year={2021},
  publisher={Institute of Mathematical Statistics}
}

@inproceedings{oettershagen2020temporal,
  title={Temporal graph kernels for classifying dissemination processes},
  author={Oettershagen, Lutz and Kriege, Nils M and Morris, Christopher and Mutzel, Petra},
  booktitle={Proceedings of the 2020 SIAM International Conference on Data Mining},
  pages={496--504},
  year={2020},
  organization={SIAM}
}

@article{david2020networkdata,
  title={Networkdata: repository of network datasets},
  author={David, Schoch},
  journal={R Package},
  year={2020}
}

@article{stanley2016clustering,
  title={Clustering network layers with the strata multilayer stochastic block model},
  author={Stanley, Natalie and Shai, Saray and Taylor, Dane and Mucha, Peter J},
  journal={IEEE transactions on network science and engineering},
  volume={3},
  number={2},
  pages={95--105},
  year={2016},
  publisher={Ieee}
}

@article{reyes2016stochastic,
  title={Stochastic blockmodels for exchangeable collections of networks},
  author={Reyes, Perla and Rodriguez, Abel},
  journal={arXiv preprint arXiv:1606.05277},
  year={2016}
}

@article{diquigiovanni2019analysis,
  title={Analysis of association football playing styles: An innovative method to cluster networks},
  author={Diquigiovanni, Jacopo and Scarpa, Bruno},
  journal={Statistical modelling},
  volume={19},
  number={1},
  pages={28--54},
  year={2019},
  publisher={SAGE Publications Sage India: New Delhi, India}
}

@article{camerlenghi2019latent,
  title={Latent nested nonparametric priors (with discussion)},
  author={Camerlenghi, Federico and Dunson, David B and Lijoi, Antonio and Pr{\"u}nster, Igor and Rodr{i}guez, Abel},
  journal={Bayesian Analysis},
  volume={14},
  number={4},
  pages={1303},
  year={2019},
  publisher={NIH Public Access}
}

@article{dahl2022search,
  title={Search algorithms and loss functions for Bayesian clustering},
  author={Dahl, David B and Johnson, Devin J and M{\"u}ller, Peter},
  journal={Journal of Computational and Graphical Statistics},
  volume={31},
  number={4},
  pages={1189--1201},
  year={2022},
  publisher={Taylor \& Francis}
}

\newpage
\appendix
\newcommand\rp{{r'}}
\newcommand\mbar{\bar m}
\newcommand\sn{{\omega}}

\section{Efficient implementation of the samplers}
\label{sec:efficiency}

Here, we collect comments on some tricks to optimize the implementation of the samplers, with pointers to the relevant functions in the code.

\subsection{Efficient updating of the block sums}
To simplify, consider a single-layer SBM with an $n \times n$ symmetric adjacency matrix $\Ab = (A_{st})$ and label vector $\xib = (\xi_s)$.
Assume that $A_{ss} = 0$ for all $s$. Let
\begin{align*}
	m_{xy} = \sum_{(s,t) \in \Gamma_{xy}} A_{st} \ind{\xi_s = x, \,\xi_t = y}
\end{align*}
and let $m_{xy}'$ be the same sum where we change a single coordinate of $\xib$, say $\xi_\sn$ to $\xi'_{\sn}$. Here
$\Gamma_{xy} := \{(s,t) :\; 1\le s \neq t \le n_j \}$ if $x \neq y$ and $\Gamma_{xx} := \{(s,t) :\; 1\le s < t \le n_j \}$. Considering the case $x\neq y$, we have
\begin{align*}
	m_{xy}' - m_{xy} &= \sum_{t \neq \sn} A_{\sn t} (\ind{\xi'_\sn = x} - \ind{\xi_\sn = x}) \ind{\xi_t = y} \\ 
 &\qquad+ \sum_{s \neq \sn} A_{s\sn} (\ind{\xi'_\sn = y} - \ind{\xi_\sn = y}) \ind{\xi_s = x} 
\end{align*}

Let $\delta_{x} := \ind{\xi'_\sn = x} - \ind{\xi_\sn = x}$ and
$
U_{y} = \sum_{t \neq \sn} A_{\sn t} \ind{\xi_t = y}.
$
By symmetry
\begin{align*}
	m'_{xy} - m_{xy} = \delta_{x} U_{y} +  \delta_{y} U_{x}, \quad x \neq y
\end{align*}
For $x= y$, 
\begin{align*}
	m_{xx}' - m_{xx} &= \sum_{\sn < t} A_{\sn t} (\ind{\xi'_\sn = x} - \ind{\xi_\sn = x}) \ind{\xi_t = x} \\ &\qquad +
	\sum_{s < \sn} A_{s\sn} (\ind{\xi'_\sn = x} - \ind{\xi_\sn = x}) \ind{\xi_s = x}  \\
	&= \sum_{t: \, t \neq \omega } A_{\sn t} (\ind{\xi'_\sn = x} - \ind{\xi_\sn = x}) \ind{\xi_t = x} = \delta_x U_x
\end{align*}
where the second line is by symmetry of $\Ab$. 
The rule for computing $D = m' - m$ is
\begin{align*}
	D &\gets \delta U^T  + U^T \delta \\
	D_{xx} &\gets D_{xx}/2, \quad \forall x
\end{align*}
This operation is implemented in  \mintinline{cpp}{comp_blk_sums_diff_v1()} in the code. Now, let 
\begin{align*}
	M_{xy} = \sum_{(s,t) \in \Gamma_{xy}} \ind{\xi_s = x, \,\xi_t = y}
\end{align*}
and $M'_{xy}$ be similarly based on changing $\xi_\sn$ to $\xi_{\sn'}$. By the same argument, letting $V_{y} = \sum_{t \neq \sn} \ind{\xi_t = y}$, the same updates apply to $\Delta = M' - M$ with $U$ replaced by $V$. Since $\mbar = M - m$, we have the following
\begin{align*}
	m' &\gets m + D \\
	\mbar' &\gets \mbar + \Delta - D 
\end{align*}

\subsection{Efficient sampling of z.}

Recall that the collapsed joint distribution for Mult-SBM is
\begin{align*}
	p(\Ab, z, \pi) \propto \prod_{x \le y} B(m_{x,y} + \alpha, \mbar_{x,y} + \beta) \prod_{i=1}^n \pi_{z_i}.
\end{align*} 
Assume that we want to sample from $p(z_s = \rp \mid z_{-s}), \; \rp \in [K]$.

Let $z_s = r$ be the previous label value. Let $m = (m_{xy})$ and $m' = (m'_{xy})$ represent the old and new values of the $m$-matrix, and similarly for $\mbar$ and $\mbar'$. The only difference between the two is that we change $z_s$ from $r$ to $r'$. We have
\begin{align*}
	\frac{p(z_s = r' \mid z_{-s})}{p(z_s = r \mid z_{-s})} 
	&=\frac{\pi_{r'}}{\pi_r}   \prod_{x \le y}
 \frac{B(m_{x,y}' + \alpha, \mbar_{x,y}' + \beta)}{B(m_{x,y} + \alpha, \mbar_{x,y} + \beta)} =:
 \frac{\pi_{r'}}{\pi_r}   \prod_{x \le y} f(x,y)
\end{align*}
Note that $\prod_{x \le y} f(x,y)$ above takes $O(K^2)$ operations and since we are doing it for $r' \in [K]$, the total cost is $O(K^3)$. However, since $m'$ is different from $m$ only over rows and columns indexed by $\{r,\rp\}$, $f(x,y) \neq 1$ only over those rows and columns and $f(x,y) = 1$ otherwise, hence the product can be computed in $O(K)$ for every $\rp$, reducing the total cost to $O(K^2)$. 

Iterating over $r' \in [K]$ requires us to run \mintinline{cpp}{comp_blk_sums_diff_v1()} for all the $K$ possibilities to compute $m'$ and $\mbar'$ each time. Note that $U$ and $V$ are only computed once and passed to \mintinline{cpp}{comp_blk_sums_diff_v1()}. We are \emph{not} recomputing $U$ and $V$ for each $r'$.

\subsection{Product over a symmetric function}

\begin{figure}[t]
    \includegraphics[width=\textwidth]{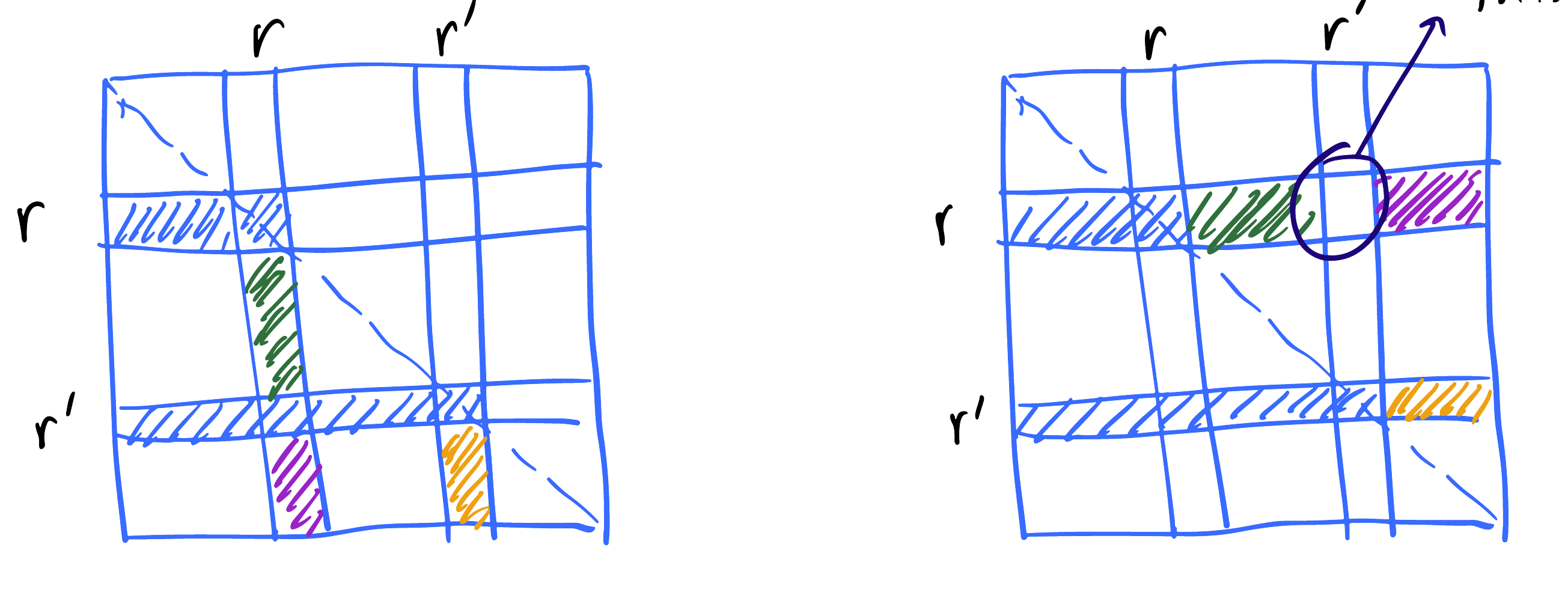}
	\caption{Efficient computation of $\prod_{x\le y} f(x,y)$. Using symmetry of $f$, we can ``almost'' just multiply elements over two rows $r$ and $\rp$. Only one element is double-counted which can be divided by.}
	\label{fig:sym:prod}
\end{figure}

Figure~\ref{fig:sym:prod} shows that we can simplify the computation further, leveraging the symmetry of $f$. We just multiply over rows $r$ and $\rp$ and discount the double-counting of $f(r,\rp)$:
\begin{align*}
	\frac{p(z_s = r' \mid z_{-s})}{p(z_s = r \mid z_{-s})} 
	&=\frac{\pi_{r'}}{\pi_r}  \prod_{y \neq \rp} f(r,y) \prod_{y} f(\rp,y).
\end{align*} 
The product calculation is implemented as \mintinline{cpp}{sym_prod()} in the code.

\newcommand\dbar{\bar d}
\subsection{Optimizing Beta ratio calculation}

In the collapsed sampler, we have to compute ratios of Beta functions of the form
\begin{align*}
	I = \frac{B(\alpha + d, \beta + \dbar)}{B(\alpha, \beta)}
\end{align*}
where $\alpha, \beta \in \reals_+$ and $d, \dbar \in \ints$ with $\alpha + d > 0$ and $\beta + \dbar > 0$. For sparse networks, both the numerator and denominator are close to zero, making the computation of $I$ challenging if we rely on standard Beta calculation procedures.  

Instead, we note that
\begin{align*}
	I =  \frac{\Gamma(\alpha + d) \Gamma(\beta + \dbar)}{\Gamma(\alpha + \beta + d + \dbar)} \frac{\Gamma(\alpha + \beta)}{\Gamma(\alpha) \Gamma(\beta)}
	= \frac{\Gamma(\alpha + d)}{\Gamma(\alpha)} \frac{\Gamma(\beta + \dbar)}{\Gamma(\beta)} 
	\Bigl[ \frac{\Gamma(\alpha + \beta + d + \dbar)}{\Gamma(\alpha + \beta)}\Bigr]^{-1}
\end{align*}
when $d > 0$, and we have 
\begin{align*}
	\Gamma(d + \alpha) / \Gamma(\alpha) = \alpha^{(d)}
\end{align*}
where $x^{(d)} = x(x+1) \cdots(x+d-1)$ is the rising factorial. The case $d < 0$ can be reduced to the previous case by inverting the ratio. Computing the log of these Gamma ratios via the rising factorial formula is a robust numerical calculation. 

\end{document}